\documentclass[notitlepage,superscriptaddress,nofootinbib,tightenlines,preprintnumbers,twocolumn]{revtex4-2}
\usepackage{amsmath,verbatim,latexsym,amssymb,indentfirst,mathrsfs,mathtools,amsthm,bbm,bm,hyperref,url,cancel,enumitem,soul}
\usepackage[font=small,labelfont=bf,format=plain,justification=raggedright,singlelinecheck=false]{caption}
\usepackage{verbatim,indentfirst}
\usepackage[title,titletoc]{appendix}

\usepackage{graphicx}
\usepackage{epstopdf}
\newcommand{\caphead}[1]{{\bf #1}}

\renewcommand{\thesection}{\Roman{section}}
\renewcommand{\thesubsection}{\Roman{section} \Alph{subsection}}
\renewcommand{\thesubsubsection}{\Roman{section} \Alph{subsection} \arabic{subsubsection}}
\makeatletter
\def\p@subsection{}
\makeatother
\makeatletter
\def\p@subsubsection{}
\makeatother

\usepackage{IEEEtrantools,physics}

\newcommand{\as}{{\tt A}} % Denotes subsystem A
\newcommand{\bs}{{\tt B}} % Denotes subsystem B
\newcommand{\xs}{{\tt X}} % Denotes generic subsystem

\newcommand{\id}{\mathbbm{1}}

\newcommand{\schrg}{\sigma_\text{chrg}}
\newcommand{\ssurp}{\sigma_\text{surp}}
\newcommand{\straj}{\sigma_\text{traj}}

\def\id{\mathbbm{1}}   % Identity

\newcommand{\GGE}{{\rm GGE}}
\newcommand{\LParen}{ \bm{(} }
\newcommand{\RParen}{ \bm{)} }
\newcommand{\dummy}{\omega} % Dummy quantum state

\renewcommand\th{ {\rm th} }

\theoremstyle{definition}

\usepackage{color}
\usepackage[dvipsnames]{xcolor}
\usepackage[normalem]{ulem}

\begin{abstract}
    We extend entropy production to a deeply quantum regime involving noncommuting conserved quantities. Consider a unitary transporting conserved quantities (``charges'') between two systems initialized in thermal states. Three common formulae model the entropy produced. They respectively cast entropy as an extensive thermodynamic variable, as an information-theoretic uncertainty measure, and as a quantifier of irreversibility. Often, the charges are assumed to commute with each other (e.g., energy and particle number). Yet quantum charges can fail to commute. Noncommutation invites generalizations, which we posit and justify, of the three formulae. Charges’ noncommutation, we find, breaks the formulae’s equivalence. Furthermore, different formulae quantify different physical effects of charges’ noncommutation on entropy production. For instance, entropy production can signal contextuality---true nonclassicality---by becoming nonreal. This work opens up stochastic thermodynamics to noncommuting---and so particularly quantum---charges.
\end{abstract}
\begin{document}

\title{Non-Abelian transport distinguishes three usually equivalent notions of entropy production}

\author{Twesh~Upadhyaya}
\email{tweshu@umd.edu}
\thanks{The first two authors contributed equally.}
\affiliation{Joint Center for Quantum Information and Computer Science, NIST and University of Maryland, College Park, MD 20742, USA}
\affiliation{Department of Physics,
University of Maryland, College Park, MD 20742, USA}
\author{William~F.~Braasch,~Jr.}
\email{wbraasch@umd.edu}
\thanks{The first two authors contributed equally.}
\affiliation{Joint Center for Quantum Information and Computer Science, NIST and University of Maryland, College Park, MD 20742, USA}
\author{Gabriel~T.~Landi}
\email{glandi@ur.rochester.edu}
\affiliation{Department of Physics and Astronomy, University of Rochester, Rochester, New York 14627, USA}
\author{Nicole~Yunger~Halpern}
\email{nicoleyh@umd.edu}
\affiliation{Joint Center for Quantum Information and Computer Science, NIST and University of Maryland, College Park, MD 20742, USA}
\affiliation{Institute for Physical Science and Technology, University of Maryland, College Park, MD 20742, USA}

\date{\normalsize\today} 

\maketitle

\section{Introduction}
\label{sec_intro}

Thermodynamics describes the transport of energy and other quantities between systems that have equilibrated individually but not with each other. These quantities are conserved according to principles such as the first law of thermodynamics.
The second law decrees which 
transport can and cannot occur spontaneously. 
Over the last three decades, researchers have revolutionized the study of transport within microscopic systems,
where fluctuations dominate~\cite{evans_probability_1993,Gallavotti_95_Dynamical,crooks_nonequilibrium_1998,kurchan2001quantum,Tasaki_00_Jarzynski, Jarzynski2004, Campisi_11_Colloquium}.

More-recent results have prompted questions about how such exchanges occur in quantum systems, which can have coherences and nonclassical correlations~\cite{Jennings2010,Alhambra2016,aberg2018, kwon2019,Levy2020, micadei2020quantum, micadei2021experimental, rodrigues2023nonequilibrium}.
To infer about quantum currents, one must measure quantum systems. But measurement back-action can destroy coherences.
Thermodynamic processes therefore depend on our measurements of them. This conundrum has inspired considerable research, but no resolution is universally agreed upon.

Quantum coherences result from the noncommutation of operators. In classical thermodynamics, a system's energy and particle number are commuting quantities; they can be measured simultaneously. What if the globally conserved thermodynamic quantities (\emph{charges}) fail to commute?

This recently posed question has upended intuitions and engendered a burgeoning subfield of quantum thermodynamics~\cite{jaynes1957information,balian1987equiprobability,vaccaro2011information,lostaglio2014resource,nyh2018beyond,lostaglio2017thermodynamic,guryanova2016thermodynamics,nyh2016microcanonical,hinds2018quantum,mitsuhashi2022characterizing,manzano2018squeezed,nyh2020noncommuting,Manzano2022,shahidani2022thermodynamic,zhang2020stationary,marvian2022restrictions,marvian2021qudit,marvian2022rotationally,Marvian_23_Non,Murthy_23_Non,Noh_23_Eigenstate,Majidy_23_Non,ito2018optimal,gour2018quantum,marvian2008building,popescu2020reference,bera2019quantum,sparaciari2020first,khanian2020resource,khanian2020quantum,fukai2020noncommutative,NYH_22_How,Kranzl_23_Experimental,majidy2023critical}; see~\cite{noncommrev23} for a recent Perspective.
For example, charges' noncommutation hinders arguments for the thermal state's form \cite{nyh2016microcanonical,nyh2018beyond} and alters the eigenstate thermalization hypothesis, which explains how quantum many-body systems thermalize internally~\cite{Murthy_23_Non}.
Other results span resource theories~\cite{nyh2018beyond,guryanova2016thermodynamics,lostaglio2017thermodynamic,nyh2016microcanonical,gour2018quantum,sparaciari2020first,marvian2008building,popescu2020reference,khanian2020resource,khanian2020quantum}, heat engines~\cite{ito2018optimal}, metrology~\cite{LPGP_24_Estimation} and a trapped-ion experiment~\cite{Kranzl_23_Experimental,nyh2020noncommuting}.
In the linear-response regime, noncommuting charges reduce entropy production ~\cite{Manzano2022}. 

We analyze entropy produced by arbitrarily far-from-equilibrium exchanges of noncommuting charges. Three formulae for entropy production, which equal each other when all charges commute, have been used widely \cite{Landi2021}. We show that these formulae fail to equal each other when charges fail to commute. This incommensurability stems from measurement disturbance: Noncommuting charges' currents cannot be measured simultaneously.

\begin{figure}[b]
\centering
\includegraphics[width=0.40\textwidth]{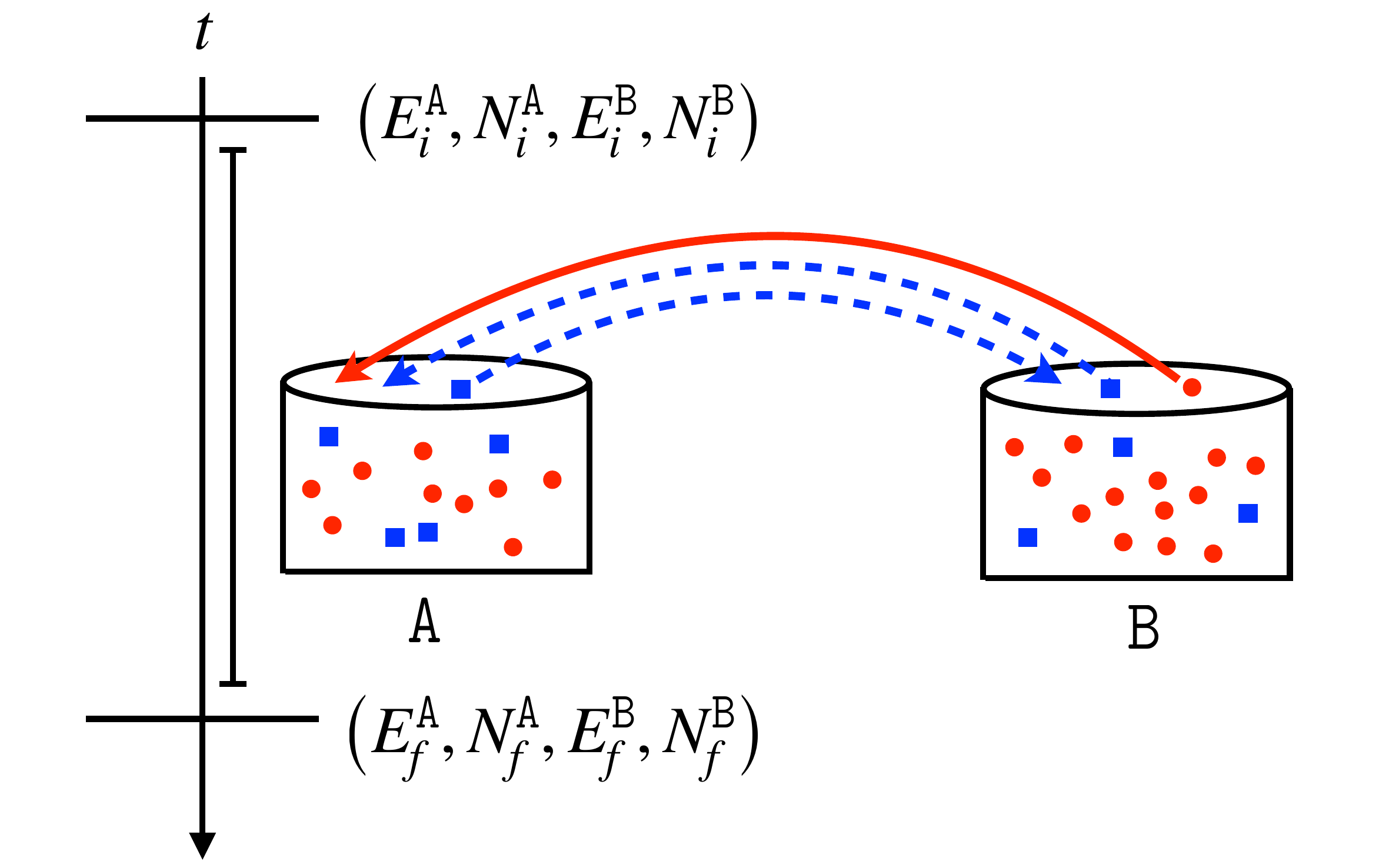}
\caption{\caphead{Two thermal systems exchanging charges:} In a paradigm ubiquitous across thermodynamics, two systems locally exchange charges that are conserved globally. The charge flow produces entropy. During each realization of the process, each system begins with some amount of every charge (e.g., energy and particle number) and ends with some amount. These amounts define a stochastic trajectory.}
\label{fig_Interaction_Traj}
\end{figure}

An example illustrates the key idea:
Consider two classical thermodynamic observables, such as energy and particle number. Let classical systems $\as$ and $\bs$ begin in thermal (grand canonical) ensembles:
System ${\tt X} = \as, \bs$ has energy $E^{\tt X}$ and particle number $N^{\tt X}$ with a probability 
$\propto e^{- \beta^{\tt X} ( E^{\tt X} - \mu^{\tt X} N^{\tt X} ) }$.
$\beta^{\tt X}$ denotes an inverse temperature; and 
$\mu^{\tt X}$, a chemical potential.
An interaction can shuttle energy and particles between the systems.
If conserved globally, the quantities are called \emph{charges}. Any charge (e.g., particle) entering or leaving a system produces entropy. The entropy produced in a trial---the \emph{stochastic entropy production} (SEP)---is a random variable. Its average over trials is non-negative, according to the second law of thermodynamics.
The SEP obeys constraints called \emph{exchange fluctuation theorems}---tightenings of the second law of thermodynamics~\cite{Jarzynski2004}.
Classical fluctuation theorems stem from a probabilistic model: System $\as$ begins with energy $E^{\as}_i$ and with $N^{\as}_i$ particles, $\as$ ends with energy $E^{\as}_f$ and with $N^{\as}_f$ particles, and $\bs$ satisfies analogous conditions, with a joint probability
$p ( E^{\as}_i, N^{\as}_i, E^{\bs}_i, N^{\bs}_i \, ; \, E^{\as}_f, N^{\as}_f, E^{\bs}_f, N^{\bs}_f )$.
The progression 
$( E^{\as}_i, N^{\as}_i, E^{\bs}_i, N^{\bs}_i )$ $\mapsto$
$( E^{\as}_f, N^{\as}_f, E^{\bs}_f, N^{\bs}_f )$, we view as a two-step \emph{stochastic trajectory} between microstates (Fig.~\ref{fig_Interaction_Traj}).

In the commonest quantum analog, quantum systems 
$\xs = \as, \bs$ begin in reduced grand canonical states 
$\propto e^{ - \beta^{\tt X} ( \hat{H}^{\tt X} - \mu^{\tt X} \hat{N}^{\tt X} ) }$. $\hat{H}^{\tt X}$ denotes a Hamiltonian; and $\hat{N}^{\tt X}$, a particle-number operator.
In the \emph{two-point measurement scheme}~\cite{kurchan2001quantum, Tasaki_00_Jarzynski},
one strongly measures each system's Hamiltonian and particle number.
Then, a unitary couples the systems, conserving 
$\hat{H}^{\as} + \hat{H}^{\bs}$ and $\hat{N}^{\as} + \hat{N}^{\bs}$.
Finally, one measures $\hat{H}^{\as}$, $\hat{H}^{\bs}$, $\hat{N}^{\as}$, and $\hat{N}^{\bs}$ again.\footnote{
Wherever necessary, to ensure that operators act on the appropriate Hilbert space, we implicitly pad with identity operators.
For example, $\hat{H}^{\as} + \hat{H}^{\bs}
\equiv \hat{H}^{\as} \otimes \hat{\id}^{\bs} + \hat{\id}^{\as} \otimes \hat{H}^{\bs}$.}
A joint probability distribution governs the four measurement outcomes. Using the outcomes and distribution, one can similarly define SEP, prove fluctuation theorems, and define stochastic trajectories~\cite{Jarzynski2004}.
Each measurement may disturb the quantum system, however~\cite{Solinas_15_Full,Miller_17_Time,Levy2020}.

Noncommutation introduces a twist into this story. Let the quantum systems above exchange charges that fail to commute with each other.
For example, consider qubits $\xs=\as,\bs$ exchanging spin components $\hat{\sigma}_{x,y,z}^{\tt X}$.
The corresponding thermal states
are $\propto \exp ( - \sum_{\alpha = x,y,z} \beta_\alpha^{\tt X} \hat{\sigma}_\alpha^{\tt X} )$,
wherein $\beta_\alpha^{\tt X}$ denotes a generalized inverse temperature~\cite{nyh2016microcanonical,nyh2020noncommuting,Kranzl_23_Experimental,Murthy_23_Non}. One cannot implement the two-point measurement scheme straightforwardly, as no system's $\hat{\sigma}_{x,y,z}^{\tt X}$ operators can be measured simultaneously. One can measure each qubit's three spin components sequentially, couple the systems with a charge-conserving unitary, and measure each qubit's $\hat{\sigma}_{x,y,z}$ sequentially again. Yet these measurements wreak havoc worse than if the charges commute: They disturb not only the systems' states, but also the subsequent, noncommuting measurements~\cite{Leggett_85_Quantum}.

\emph{Weak measurements} would disturb the systems less~\cite{allahverdyan_nonequilibrium_2014, Lostaglio_22_Kirkwood}
at the price of extracting less information~\cite{Dressel2014}.
As probabilities describe strong-measurement experiments, \emph{quasiprobabilities} describe weak-measurement experiments (App.~\ref{app_Motivate_KDQ}). Quasiprobabilities resemble probabilities---being normalized to 1, for example. They violate axioms of probability theory, however, as by becoming negative. 
Consider, then, replacing the above protocol's strong measurements with weak measurements. We may loosely regard $\as \bs$ as undergoing stochastic trajectories weighted by quasiprobabilities, rather than probabilities. Levy and Lostaglio applied quasiprobabilities in deriving a fluctuation theorem for energy exchanges~\cite{Levy2020}.
Their fluctuation theorem contains the real part of a \emph{Kirkwood--Dirac quasiprobability} (KDQ)~\cite{kirkwood_quantum_1933, dirac_on_1945,Lostaglio_22_Kirkwood,Gherardini_24_Quasi,DRMAS_24_Properties}.

KDQs have recently proven useful across quantum thermodynamics~\cite{allahverdyan_nonequilibrium_2014,NYH_17_Jarzynski,Lostaglio_18_Quantum,Lostaglio_20_Certifying, Levy2020, Lostaglio_22_Kirkwood,Santini_23_Work}, information scrambling~\cite{NYH_17_Jarzynski, NYH_18_Quasiprobability,NYH_19_Entropic,JRGA_19_Out,Mohseninia_19_Optimizing}, tomography~\cite{Lundeen_11_Direct,Lundeen_12_Procedure,Hofmann_14_Sequential,Salvail_13_Full,Malik_14_Direct,Howland_14_Compressive,Mirhosseini_14_Compressive}, metrology~\cite{Arvidsson-Shukur2020,Lupu_22_Negative,Jenne_22_Unbounded}, and foundations~\cite{Kunjwal_19_Anomalous,DRMAS_21_Conditions,DeBievre_21_Complete,DeBievre_23_Relating,Gao_23_Measuring,Rastegin_23_On}. 

When employed as work or heat distributions, KDQs have multiple desirable properties not achieved by joint probability distributions~\cite{allahverdyan_nonequilibrium_2014, Lostaglio_22_Kirkwood, francica2022a,francica2022b}

Negative and nonreal KDQs can reflect nonclassicality~\cite{Pusey2014, Kunjwal2019} and measurement disturbance~\cite{Jozsa_07_Complex,Hofmann_11_On,Hofmann_12_Complex,Dressel_12_Significance,Hofmann_14_Derivation,Hofmann_14_Sequential}.
In~\cite{Levy2020}, negative real KDQs signal anomalous heat currents, which flow spontaneously from a colder to a hotter system.
Generalizing from energy to potentially noncommuting charges, our results cover a more fully quantum setting. They also leverage the KDQ's ability to become nonreal.

To accommodate noncommuting charges, we generalize the SEP.
In conventional quantum thermodynamics, three common SEP formulae equal each other~\cite{Landi2021}. Entropy is cast as an extensive thermodynamic variable by a ``charge formula,'' as quantifying missing information 
by a ``surprisal formula,'' and as quantifying irreversibility by a ``trajectory formula.''
We generalize all three formulae using KDQs.
If the charges commute, the generalizations reduce to the usual formulae.
The generalizations  
satisfy four sanity checks, including by equaling each other when the charges commute.
Yet we find deductively that noncommuting charges break the equivalence, generating three species of SEP.

Different SEPs, we find, highlight different ways in which charges' noncommutation impacts transport:
\begin{enumerate}

   \item \emph{Charge SEP:} Charges' noncommutation enables individual stochastic trajectories to violate charge conservation. These violations underlie commutator-dependent corrections to a fluctuation theorem.

   \item \emph{Surprisal SEP:} Initial coherences, relative to eigenbases of the charges, enable the average surprisal entropy production to become negative. Such negativity simulates a reversal of time's arrow. Across thermodynamics, a ubiquitous initial state is a product of thermal states. Only if the charges fail to commute can such a state have the necessary coherences. Hence charges' noncommutation enables a resource, similar to work, for effectively reversing time's arrow in a common setup.
    
   \item \emph{Trajectory SEP:} The generalized trajectory SEP can become nonreal, thanks to the KDQ. Such nonreality signals contextuality---a strong form of nonclassicality~\cite{Bell_66_On, Kochen_67_The, Spekkens2005}---in a noncommuting-current experiment. 
   
\end{enumerate}
This work opens the field of stochastic thermodynamics~\cite{Strasberg_22_Quantum} to noncommuting charges. Furthermore, our work advances the widespread research program of critically comparing thermodynamic with information-theoretic entropies and leveraging information theory for thermodynamics \cite{bennett82,Lloyd_88_Black,maxwellbook90,
Parrondo_15_Thermodynamics}. 

The paper is organized as follows.
Section~\ref{prelim} details the setup and reviews background material. Section~\ref{sec_3formulae} presents the three SEP formulae, the physical insights they imply, and their fluctuation theorems and averages. 
Section~\ref{example} numerically illustrates our main results with a two-qubit example. In Sec.~\ref{sec_experiment}, we sketch a trapped-ion experiment based on our results. Section~\ref{conc} concludes with avenues for future work.

As a semantic disclaimer and to provide broader context: Studies of the second law have engendered debates about what should be called \emph{the} entropy production.
We provide an alternative perspective: We show that three SEP formulae, despite being equivalent in the commuting case, become inequivalent if charges cease to commute.
Each of these formulae encapsulates a different aspect of entropy production.
This perspective harmonizes with demonstrations that, for small systems, the conventional second law splits into multiple second ``laws''~\cite{Horodecki_13_Fundamental,Brandao2015second,Lostaglio_15_Description,NYH_16_Beyond,nyh2018beyond,nyh2016microcanonical,Gour_18_Quantum}.

\section{Preliminaries}
\label{prelim}

We specify the physical setup in Sec.~\ref{bgsetup}. Section~\ref{KD} reviews KDQs;
Sec.~\ref{entr}, information-theoretic entropies;
and Sec.~\ref{sec_FTs}, fluctuation theorems.

\subsection{Setup}
\label{bgsetup}

Consider two identical quantum systems, $\as$ and $\bs$. (From now on, we omit hats from operators.) Each system corresponds to a copy of a Hilbert space $\mathcal{H}$.
The bipartite initial state $\rho$ leads to the reduced states
$\rho^{\as} \coloneqq \Tr_{\bs}(\rho)$ and
$\rho^{\bs}\coloneqq \Tr_{\as}(\rho)$.

On $\mathcal{H}$ are defined single-system observables
$Q_{\alpha = 1, 2, \ldots, c}$, assumed to be linearly independent~\cite{NYH_22_How}.
For $\xs = \as, \bs$, we denote $\xs$'s copy of $Q_\alpha$ by $Q_\alpha^{\tt X}$.
The dynamics will conserve the global observables $Q_\alpha^\text{tot}\coloneqq Q_\alpha^\as \otimes \id^\bs + \id^\as \otimes Q_\alpha^\bs$,
so we call the $Q_\alpha$'s charges.
 In the \textit{commuting case}, all the charges commute: $[Q_\alpha, Q_{\alpha'}] = 0$ $\forall \alpha, \alpha'$.
At least two charges do not in the \textit{noncommuting case}: $[Q_\alpha, Q_{\alpha'}] \neq 0$ for some $\alpha, \alpha'$.
Each charge eigendecomposes as 
$Q_\alpha = \sum_j \lambda_{\alpha,j} \, \Pi_{\alpha, j}$. 
We denote tensor-product projectors by
$\Pi_{\alpha, {k}} 
= \Pi^{\as}_{\alpha, k^\as} \otimes \Pi^{\bs}_{\alpha,k^\bs}$, invoking the composite index ${k}=(k^{\as},k^{\bs})$.
To simplify the formalism, we assume that the $Q_\alpha$ are nondegenerate, as in~\cite{Jarzynski2004, Esposito_2009_Nonequilibrium, Jevtic_15_Exchange, Levy2020}.
Appendix~\ref{app_degen} concerns extensions to degenerate charges.
The nondegeneracy renders every projector rank-one: 
$\Pi_{\alpha, k} =\ketbra{\alpha_k}{\alpha_k}$.
We call the $Q_\alpha^\text{tot}$ eigenbasis
$\{\ket{\alpha_k}\}_k$ the \textit{$\alpha^\text{th}$ product basis}.

Having introduced the charges, we can expound upon the initial state. The reduced states $\rho^\xs$ are \emph{generalized Gibbs ensembles} (GGEs)~\cite{Rigol_09_Breakdown, Rigol_07_Relaxation, Langen_2015_Experimental, Vidmar_2016_Generalized, Essler_2016_Quench}
\begin{align}
   \label{eq_GGE}
   \rho^{\tt X}_\GGE
   \coloneqq \frac{1}{Z^{\tt X}} \exp \Bigg( - \sum_{\alpha = 1}^c  
   \beta^{\tt X}_\alpha Q_\alpha^{\tt X}  \Bigg) .
\end{align}
$\beta^{\tt X}_\alpha$ denotes a generalized inverse temperature.
The partition function $Z^{\tt X}$ normalizes the state.
If the $Q_\alpha^{\tt X}$'s fail to commute, the GGE is often called the \emph{non-Abelian thermal state}~\cite{nyh2016microcanonical,nyh2020noncommuting,Kranzl_23_Experimental,Murthy_23_Non}.
Across most of this paper, $\rho$ equals the tensor product
\begin{equation}\label{initstate}
    \rho^{\as}_\GGE \otimes \rho^\bs_\GGE.
\end{equation}
Such a product of thermal states, being a simple nonequilibrium state, surfaces across thermodynamics~\cite[Sec.~3.6]{Vinjanampathy_16_Quantum}.
(In Sections~\ref{sec_surp_avg} and~\ref{sec_traj_avg}, we consider initial states $\rho$ that deviate from this form but retain GGE reduced states.)
 
Our results stem from the following protocol: Prepare $\as \bs$ in $\rho$. Evolve $\rho$ under a unitary $U$ that conserves every global charge: 
\begin{align}
\label{chargecons}
   [U, Q_\alpha^\text{tot}]=0
   \; \forall \alpha .
\end{align}
$U$ can bring the state arbitrarily far from equilibrium.
The final state, $\rho_\text{f} = U \rho U^\dagger$,
induces the reduced states
$\rho_\text{f}^\as \coloneqq \Tr_\bs ( \rho_\text{f} )$ and
$\rho_\text{f}^\bs \coloneqq \Tr_\as ( \rho_\text{f} )$.
The average amount of $\alpha$-type charge in $\as$ changes by
$\Delta \langle Q_\alpha \rangle 
\coloneqq \Tr( Q_\alpha^\as [\rho^\as_\text{f} - \rho^\as] )$; and the average amount in $\bs$, by
$\Tr( Q_\alpha^\bs [\rho^\bs_\text{f} - \rho^\bs] )
= - \Delta \langle Q_\alpha \rangle$.
The systems' generalized inverse temperatures differ by
$\Delta \beta_\alpha \coloneqq \beta_\alpha^\as - \beta_\alpha^\bs$.

\subsection{Kirkwood-Dirac quasiprobability}
\label{KD}

The KDQ's relevance stems from the desire to reason about charges flowing between $\as$ and $\bs$. One can ascribe to $\as$ some amount of $\alpha$-type charge only upon measuring $Q_\alpha^\as$. Measuring $Q_\alpha^\as$ strongly would disturb $\as$ and subsequent measurements of noncommuting $Q_{\alpha'}^\as$'s~\cite{Leggett_85_Quantum}. Therefore, we consider sequentially measuring charges weakly.
We define the \emph{forward protocol} by combining the preparation procedure and unitary with weak measurements (Sec.~\ref{traj} introduces a reverse protocol): 
\begin{enumerate}
   \item Prepare $\as \bs$ in $\rho$. 
   \item Weakly measure the product basis of $Q_1^\as$ and $Q_1^\bs$, then the product basis of of $Q_2^\as$ and $Q_2^\bs$, and so on until $Q_c^\as$ and $Q_c^\bs$. (One can implement these measurements using the detector-coupling technique in~\cite[footnote~9]{NYH_18_Quasiprobability}.) 
   
   \item \label{step_U}
   Evolve $\as \bs$ under $U$. 
   \item \label{step_final_meas}
   Weakly measure the charges' product bases in the reverse order, from $c$ to 1. (This measurement ordering ensures that our SEP definitions satisfy sanity checks described in Sec.~\ref{sec_3formulae}.)
\end{enumerate}
Lacking strong measurements, this protocol differs qualitatively from two-point-measurement schemes. 
The forward protocol leads naturally to a KDQ, as shown in App.~\ref{app_Motivate_KDQ}:
\begin{align}
   \label{KDdef} &
   \Tr \left( U^\dag
   \left[ \Pi_{1, f_1} \Pi_{2, f_2} \ldots \Pi_{c, f_c} \right]
   U 
   \left[ \Pi_{c, i_c} \ldots \Pi_{2,i_2} \Pi_{1,i_1}  \right]
   \rho \right)
   \nonumber \\ & \quad
    \eqqcolon \tilde{p}_{\rm F} (i_1, i_2, \ldots, i_c \, ; \,
    f_c, f_{c-1}, \ldots, f_1) .
\end{align}
The list
$(i_1, i_2, \ldots, i_c \, ; \,
  f_c, f_{c-1}, \ldots, f_1)$
defines a stochastic trajectory, as in Sec.~\ref{sec_intro}. (Recall the definition, in Sec. \ref{bgsetup}, of composite-system indices.) Loosely speaking, we might view the trajectory as occurring with a joint quasiprobability $\tilde{p}_{\rm F} (i_1, i_2, \ldots, i_c \, ; \,
   f_c, f_{c-1}, \ldots, f_1)$.

% \billycomm{In progress...} Paragraph about how the KDQ emerges as the unique distribution for charge flow that satisfies certain requirements (marginal properties, convexity) \cite{Lostaglio_22_Kirkwood}. Also cite \cite{perarnau-llobet2018?}?}

$\tilde{p}_{\rm F}$ can assume negative and nonreal values.
One can infer $\tilde{p}_{\rm F}$ experimentally by performing the forward protocol many times, performing strong-measurement experiments, and processing the outcome statistics~\cite{NYH_17_Jarzynski}. 
Angle brackets $\langle . \rangle$ denote averages with respect to $\tilde{p}_{\rm F}$, unless we specify otherwise.

Two cases further elucidate $\tilde{p}_{\rm F}$ and the forward protocol: the commuting case and weak-measurement limit.
In the commuting case, if $\rho$ is diagonal with respect to the charges' shared eigenbasis,
$\tilde{p}_{\rm F}$ is a joint probability.

\subsection{Information-theoretic entropies}
\label{entr}

We invoke four entropic quantities from information theory~\cite{Watrous2018}. 
Let $X = x_1, x_2, \ldots, x_n$ denote a discrete random variable; $P = \{ p_1, p_2, \ldots, p_n \}$ and $R = \{ r_1, r_2, \ldots, r_n \}$, probability distributions over $X$; and $\dummy_1$ and $\dummy_2$, quantum states.
Suppose that $X$ evaluates to $x_j$.
The \emph{surprisal} $- \log (p_j)$ quantifies the information we learn. 
(This paper's logarithms are base-$e$.)
Averaging the surprisal yields the \emph{Shannon entropy}, $S_{\rm Sh}(P) \coloneqq - \sum_j p_j \log(p_j)$. The quantum analog is the \emph{von Neumann entropy}, 
$S_{\rm vN}(\dummy_1) 
\coloneqq - \Tr \LParen \dummy_1 \log(\dummy_1) \RParen$.

The \emph{quantum relative entropy} quantifies the distance between states:
$D(\dummy_1 || \dummy_2) 
= \Tr \LParen \dummy_1 [ \log(\dummy_1) - \log(\dummy_2)] \RParen$.
$D$ measures how effectively one can distinguish between $\dummy_1$ and $\dummy_2$, on average, in an asymmetric hypothesis test. 
$D (\dummy_1 || \dummy_2) \geq 0$ vanishes if and only if $\dummy_1 = \dummy_2$.

Analogously, the \emph{classical relative entropy} distinguishes probability distributions:
$D(P || R) = \sum_j  p_j \left[ \log(p_j) - \log(r_j) \right]$.
We will substitute KDQ distributions for $P$ and $R$ in Sec.~\ref{sec_traj_def}. The logarithms will be of complex numbers. We address branch-cut conventions and the complex logarithm's multivalued nature there.

\subsection{Exchange fluctuation theorems}
\label{sec_FTs}

An exchange fluctuation theorem governs two systems trading charges. (We will drop the \emph{exchange} from the name.) Consider two quantum systems initialized and exchanging energy as in the introduction, though without the complication of particles. Each trial has a probability $p(\sigma)$ of producing entropy $\sigma$. Denote by $\langle . \rangle_P$ averages with respect to the probability distribution. The fluctuation theorem
$\expval{ e^{-\sigma} }_P = 1$ implies the second-law-like inequality $\expval{\sigma}_P \geq 0$~\cite{Jarzynski2004}. Unlike the second law, fluctuation theorems are equalities arbitrarily far from equilibrium. More-general (e.g., correlated) initial states engender corrections: 
$\expval{ e^{-\sigma} }_P = 1 + \ldots$~\cite{Jevtic_15_Exchange,Levy2020}.

\section{Three generalized SEP formulae}
\label{sec_3formulae}

We now present and analyze the generalized SEP formulae: the charge SEP $\schrg$ (Sec.~\ref{chrg}), the surprisal SEP $\ssurp$ (Sec.~\ref{surp}), and the trajectory SEP $\straj$ (Sec.~\ref{traj}). 
To simplify notation, we suppress the indices that identify the trajectory along which an SEP is produced.
% In statements true of all the formulae, we use the notation $\sigma$.  % NYH: This statement doesn't correctly characterize a use of \sigma on the LHS of p. 6. Also, the statement is arguably unnecessary.

Each formula satisfies four sanity checks: (i) Each $\langle \sigma \rangle$ has a clear physical interpretation. 
(ii) Each $\sigma$ satisfies a fluctuation theorem (Sec.~\ref{sec_FTs}). Any corrections depend on charges' commutators.
(iii) If the charges commute, all three SEPs coincide. (iv) Suppose that no current flows: $[U,Q_\alpha^\xs]=0$ $\forall \alpha$.
As expected physically, the average entropy production vanishes: $\langle \sigma \rangle = 0$. Also, the fluctuation theorems lack corrections: 
$\langle e^{-\sigma} \rangle = 1$.

\subsection{Charge stochastic entropy production}
\label{chrg}

First, we motivate the charge SEP's definition (Sec.~\ref{sec_chrg_def}). $\schrg$ portrays entropy as an extensive thermodynamic quantity (Sec.~\ref{sec_chrg_avg}).
Furthermore, $\schrg$ satisfies a fluctuation theorem whose corrections depend on commutators of the $Q_\alpha$'s (Sec.~\ref{sec_chrg_FT}). 
Corrections arise because noncommuting charges enable individual stochastic trajectories to violate a ``microscopic,'' or ``detailed,'' notion of charge conservation. [Nevertheless, charge conservation as defined in Eq.~\eqref{chargecons} is not violated.]

\subsubsection{Charge SEP formula}
\label{sec_chrg_def}

The fundamental relation of statistical mechanics~\cite{Pathria2001} motivates $\schrg$'s definition. In this paragraph, we reuse quantum notation (Sec.~\ref{bgsetup}) to denote classical objects, for simplicity. The fundamental relation governs large, classical systems ${\tt X} = {\as}, {\bs}$ that have extensive charges ${Q}^{{\tt X}}_\alpha$ and intensive parameters ${\beta}^{{\tt X}}_\alpha$.
Let an infinitesimal interaction conserve each
${Q}^{\as}_\alpha + {Q}^{\bs}_\alpha$.
Since $d {Q}_\alpha^{\bs}
= - d{Q}_\alpha^{\as}$,
the total entropy changes by
\begin{align}
   dS^{ \as \bs } 
   & = \sum_\alpha \left( 
   \beta_\alpha^{\as}  \,
   d{Q}_\alpha^{\as} 
   + \beta_\alpha^{\bs}  \,
     d {Q}_\alpha^{\bs} \right) \\
   &= \sum_\alpha \Delta {\beta}_\alpha \, 
   d {Q}_\alpha^{\as} \,. 
   \label{ftr}
\end{align}
According to the second law of thermodynamics, $dS^{ \as \bs } \geq 0$ during spontaneous processes~\cite{Callen_85_Thermodynamics}.
We posit that $\expval{\schrg}$ should assume the form~\eqref{ftr}. 
% \billy{This formula describes an average over trials, as typical trials are average trials in the thermodynamic limit~\cite{Callen_85_Thermodynamics}.
% For the charge SEP definition, we produce an expression $\sigma_{\rm chrg}$ for which $\langle \sigma_{\rm chrg} \rangle$ has the form of~\eqref{ftr}.}
We reverse-engineer such a formula, using the charges' eigenvalues:
\begin{equation}
   \schrg 
   \coloneqq \sum_\alpha \left[ 
   \beta^\as_\alpha \left( \lambda_{\alpha, f^\as_\alpha} - \lambda_{\alpha, i^\as_\alpha} \right) 
   + \beta^\bs_\alpha \left( \lambda_{\alpha, f^\bs_\alpha} 
   - \lambda_{\alpha, i^\bs_\alpha} \right)
   \right] . \label{schrg}
\end{equation}
\subsubsection{Average of charge SEP}
\label{sec_chrg_avg}

By design, the charge SEP averages to
\begin{equation}
   \label{avgc}
   \expval{\schrg} 
   = \sum_\alpha \Delta \beta_\alpha \, \Delta \expval{Q_\alpha} .
\end{equation} 
This average is non-negative, equaling a relative entropy~\cite{Landi2021}:
\begin{equation}
\label{relentr}
\expval{\schrg} = D(\rho_{\rm f}||\rho)
\geq 0 \, .
\end{equation}
This inequality echoes the second-law statement written just below Eq.~\eqref{ftr}. 

\subsubsection{Fluctuation theorem for charge SEP}
\label{sec_chrg_FT}

$\schrg$ obeys the fluctuation theorem
\begin{align} 
   \label{noncommft}
   & \expval{e^{-\schrg}} \\ \nonumber
   &= \Tr( U^\dagger   
   \left[ e^{-\Delta \beta_1 Q_1^{\as}}  \ldots 
   e^{-\Delta \beta_c Q_c^{\as} } \right] 
   U 
   \left[ e^{\Delta \beta_c Q_c^{\as}} \ldots e^{\Delta \beta_1 Q_1^{\as}} \right] 
   \rho ) 
   \\ \nonumber & \quad \;
   + \left(  \text{$c-1$ terms dependent on
   commutators} \right) .
\end{align}
We prove the theorem and present the correction's form in App.~\ref{app:schrg_FT}. Below, we show that the right-hand side (RHS) evaluates to 1 in the commuting case. In the noncommuting case, corrections arise from two physical sources: (i) $\rho^\as$ and $\rho^\bs$ are non-Abelian thermal states. (ii) Individual stochastic trajectories can violate charge conservation.

In the commuting case, the first term in Eq.~\eqref{noncommft} equals 1. The reason is, in $\rho = \rho_\GGE^\as \otimes \rho_\GGE^\bs$,
each $\rho_\GGE^{\tt X}
\propto \exp \left( - \sum_\alpha  \beta^{\tt X}_\alpha Q_\alpha^{\tt X}  \right)
\propto \prod_\alpha \exp( -\beta^{\tt X}_\alpha Q_\alpha^{\tt X} )$.
The $\as$ exponentials cancel the $e^{\beta_\alpha^\as Q_\alpha^\as}$ factors in Eq.~\eqref{noncommft}. 
No such cancellation occurs in the noncommuting case, since $\rho_\GGE^{\tt X}
\not\propto \prod_\alpha \exp( -\beta^{\tt X}_\alpha Q_\alpha^{\tt X} )$.
The first term in Eq.~\eqref{noncommft} can therefore deviate from 1, quantifying the noncommutation of the charges in $\rho_\GGE^{\tt X}$.

The second term vanishes in the commuting case, since all commutators of charges vanish. This second term can deviate from 0 in the noncommuting case, due to \emph{nonconserving trajectories}, which we introduce now. Define a trajectory 
$(i^\as_1,i^\bs_1,i^\as_2, i^\bs_2,\ldots, i^\bs_c)$ $\mapsto$ $(f^\as_c,f^\bs_c,f^\as_{c-1},f^\bs_{c-1},\ldots,f^\bs_1$)
as \emph{conserving} if the corresponding charge eigenvalues satisfy
\begin{equation}
   \label{dcc}
\lambda_{\alpha,i^\as_\alpha}+\lambda_{\alpha,i^\bs_\alpha}
   =\lambda_{\alpha,f_\alpha^\as}
   +\lambda_{\alpha,f_\alpha^\bs} 
   \quad \forall \alpha .
\end{equation}
Any trajectory that violates this condition, we call \emph{nonconserving}. Loosely speaking, under~\eqref{dcc}, $\as \bs$ has the same amount of $\alpha$-type charge at the trajectory's start and end.

In the commuting case, $\tilde{p}_{\rm F} = 0$ 
when evaluated on nonconserving trajectories. We call this vanishing \emph{detailed charge conservation}. Earlier work has relied on detailed charge conservation~\cite{allahverdyan_nonequilibrium_2014,Levy2020}.
In the noncommuting case, $\tilde{p}_{\rm F}$ need not vanish when evaluated on nonconserving trajectories (App.~\ref{app:detailed}).
The mathematical reason is, different charges' eigenprojectors in $\tilde{p}_{\rm F}$ [Eq.~\eqref{KDdef}] can fail to commute. 
A physical interpretation is that measuring any $Q_\alpha^{\tt X}$ disturbs any later measurements of noncommuting $Q_{\alpha'}^{\tt X}$'s. Nonconserving trajectories resemble classically forbidden trajectories in the path-integral formulation of quantum mechanics.

We now show that nonconserving trajectories underlie the final term in Eq.~\eqref{noncommft}. 
Every stochastic function $g(i_1, i_2, \ldots,i_c \, ; \, f_c, f_{c-1}, \ldots, f_1)$ averages to
$\expval{g} = \expval{g}_\text{cons} + \expval{g}_\text{noncons}$.
Each term equals an average over just the conserving or nonconserving trajectories. For example, the average~\eqref{avgc} decomposes as $\expval{\schrg} = \expval{\schrg}_\text{cons} + \expval{\schrg}_\text{noncons}$. 
The second term on the RHS of Eq.~\eqref{noncommft} takes the form $\expval{g}_\text{noncons}$ (App.~\ref{ftandnoncons}).
In the commuting case, all trajectories are conserving, so the second term equals zero---as expected, since the term depends on commutators $[Q_\alpha^\as,Q_{\alpha'}^\as]$.
Therefore, the fluctuation theorem's final term (second correction) stems from violations of detailed charge conservation. The average~\eqref{avgc} contains contributions from conserving and nonconserving trajectories.

We have identified two corrections to the fluctuation theorem~\eqref{noncommft}. On the equation's RHS, the first term can deviate from 1, and the second term can deviate from 0. The first deviation originates in noncommutation of the charges in $\rho^\xs_\GGE$. The second deviation stems from nonconserving trajectories. An example illustrates the distinction between noncommuting charges' influences on initial conditions and dynamical trajectories.
Let $c = 3$, and let
$[Q_1,Q_2] \neq 0$, while $[Q_1,Q_3],[Q_2,Q_3]=0$. 
Let the noncommuting charges correspond to uniform 
$\beta$'s: 
$\beta^\as_1=\beta^\bs_1 \, ,
   \quad \text{and} \quad
   \beta^\as_2=\beta^\bs_2 \, .$
Thus, there exists no temperature gradient that directly drives noncommuting-charge currents. Accordingly, the fluctuation theorem's first term equals 1. Yet the second term---an average over nonconserving trajectories---is nonzero.

\subsection{Surprisal stochastic entropy production}
\label{surp}

$\ssurp$ casts entropy as missing information (Sec.~\ref{sec_surp_SEP}). The average $\langle \ssurp \rangle$ can be expressed in terms of relative entropies (Sec.~\ref{sec_surp_avg}), as one might expect from precedent~\cite{Landi2021}. Yet initial coherences, relative to charges' product bases, can render $\langle \ssurp \rangle$ negative. If $\rho$ is a product $\rho^\as_\GGE \otimes \rho^\bs_\GGE$---as across much of thermodynamics---$\rho$ has such coherences only if the $Q_\alpha$'s fail to commute. Furthermore, positive $\langle \ssurp \rangle$ values accompany time's arrow. Hence charges' noncommutation enables a resource, similar to work, for effecting a seeming reversal of time's arrow. Charges' noncommutation also engenders a correction to the $\ssurp$ fluctuation theorem (Sec.~\ref{sec_surp_FT}).

\subsubsection{Surprisal SEP formula}
\label{sec_surp_SEP}

Information theory motivates the surprisal SEP formula. Averaging the surprisal $-\log(p_j)$ yields
the Shannon entropy (Sec. \ref{entr}), so the surprisal is a stochastic (associated-with-one-trial) entropic quantity. A difference of two surprisals forms our $\ssurp$ formula. The probabilities follow from preparing $\rho$ and projectively measuring the $\alpha^\th$ product basis, for any $\alpha$. Outcome 
$i_\alpha \coloneqq (i_\alpha^\as, i_\alpha^\bs)$ 
obtains with a probability
$p_\alpha (i_\alpha^\as, i_\alpha^\bs)
\coloneqq \Tr( \Pi_{\alpha, i_\alpha} \rho )$;
and outcome 
$f_\alpha \coloneqq (f_\alpha^\as, f_\alpha^\bs)$,  
with a probability
$p_\alpha (f_\alpha^\as, f_\alpha^\bs)
\coloneqq \Tr( \Pi_{\alpha, f_\alpha} \rho )$.
Appendix~\ref{app_surp_motivate} shows how these probabilities generalize those in the conventional surprisal-SEP formula~\cite{Landi2021}. The surprisal SEP quantifies the information gained if we expect to observe $i_\alpha$ but we obtain $f_\alpha$:
\begin{align}
   \label{ssurp}
   \ssurp
   \coloneqq \log \left( 
   \frac{p_\alpha(i^\as_\alpha, i^\bs_\alpha) }{p_\alpha(f^\as_\alpha,f^\bs_\alpha) } \right) .
\end{align}
All results below hold for arbitrary $\alpha$. Nevertheless, App.~\ref{app_symmetrized} introduces a variation on Eq.~\eqref{ssurp}---an alternative definition that contains an average over all $\alpha$.
Appendix~\ref{app:surp_chrg_agree} confirms that $\ssurp$ reduces to $\schrg$ if the $Q_\alpha$'s commute.

\subsubsection{Average of surprisal SEP}
\label{sec_surp_avg}

$\langle \ssurp \rangle$ demonstrates that charges' noncommutation can enable a seeming reversal of time's arrow.
Time's arrow manifests in, e.g., spontaneous flows of heat from hot to cold bodies. This arrow accompanies positive average entropy production. Hence negative $\langle \sigma \rangle$'s simulate reversals of time's arrow. These simulations cost resources, such as the work traditionally used to pump heat from colder to hotter bodies.
Quantum phenomena, such as entanglement, serve as such resources, too~\cite{Partovi2008, Jennings2010}.
We identify another such resource:
initial coherences relative to charges' product bases, present in the common initial state~\eqref{eq_GGE} only if charges fail to commute.

To prove this result, we denote by $\Phi_\alpha$ the channel that dephases states $\dummy$ with respect to the $\alpha^\text{th}$ product basis: 
$\Phi_\alpha(\dummy) 
\coloneqq \sum_k 
\Pi_{\alpha,k} \, \dummy \, \Pi_{\alpha,k}$.
$\ssurp$ averages to
\begin{equation}
   \label{avgs}
   \hspace{-0.6cm} \expval{\ssurp} =   
   D \LParen \rho_{\rm f} || \Phi_\alpha \left( \rho \right) \RParen 
   - D \LParen \rho || \Phi_\alpha(\rho) \RParen 
\end{equation} 
(App.~\ref{app:surp_avg}).
The relative entropy is non-negative (Sec.~\ref{entr}). Therefore, initial coherences relative to charges' product bases can reduce $\expval{\ssurp}$. Such coherences can even render $\expval{\ssurp}$ negative. 
Since $\expval{\schrg} \geq 0$ [Eq.~\eqref{relentr}], $\ssurp$ is sensitive to the resource of coherence, while $\schrg$ is not.

This result progresses beyond three existing results. First, coherences engender a correction to a heat-exchange fluctuation theorem~\cite{Levy2020}. Those coherences are relative to an eigenbasis of the only charge in~\cite{Levy2020}, energy. If the dynamics conserve only one charge, or only commuting charges, then the thermal product 
$\rho^\as_\GGE \otimes \rho^\bs_\GGE$ [Eq.~\eqref{eq_GGE}] lacks the necessary coherences.  
$\rho^\as_\GGE \otimes \rho^\bs_\GGE$ has those coherences only if charges fail to commute.
Across thermodynamics,
$\rho^\as_\GGE \otimes \rho^\bs_\GGE$ is ubiquitous as an initial state. Hence noncommuting charges underlie a resource for effectively reversing time's arrow in a common thermodynamic setup.

Second,~\cite{Manzano2022} shows that noncommuting charges can reduce entropy production in the linear-response regime. Our dynamics $U$ can be arbitrarily far from equilibrium. 
Third, just as we attribute $\expval{\ssurp} < 0$ to initial coherences, so do~\cite{Partovi2008,Lloyd_1989_Use, Jennings2010, Jevtic_15_Exchange}
attribute negative average entropy production to initial correlations. 
Our framework recapitulates such a correlation result, incidentally: If $\rho$ encodes correlations but retains GGE marginals, $\expval{\schrg}$ [Eq.~\eqref{avgc}] shares the form of~\eqref{avgs}, to within one alteration. The decorrelated state $\rho^\as \otimes \rho^\bs$ replaces the dephased state $\Phi_\alpha (\rho)$: 
$\expval{\schrg} 
= D( \rho_{\rm f} || \rho^\as \otimes \rho^\bs ) 
- D(\rho || \rho^\as \otimes \rho^\bs )$. 
Initial correlations can therefore render $\expval{\schrg}$ negative.
Just as initial correlations can render a $\langle \sigma \rangle$ negative, $\ssurp$ reveals coherences attributable to noncommuting charges can. 
% It is not unreasonable that the surprisal SEP can be negative, while the charge SEP is always nonnegative, since different definitions of SEP are sensitive to different quantum effects.}
%\billycomm{Twesh, I commented out an edit you had made here and moved it---with some changes---up two paragraphs. Okay? If so, please delete this.}
%Perfect, thanks Billy.

\subsubsection{Fluctuation theorem for surprisal SEP}
\label{sec_surp_FT}

To formulate the fluctuation theorem, we define the coherent difference
$\Delta \rho_\alpha
\coloneqq \Phi_\alpha(\rho)^{-1} -\rho^{-1} $. 
It quantifies the coherences of $\rho$ with respect to the $\alpha^\th$ product basis. If and only if $\rho$ is diagonal with respect to this basis, $\Delta \rho_\alpha = 0$. The surprisal SEP obeys the fluctuation theorem
\begin{align}
   \label{surpft}
   \expval{e^{-\ssurp}} 
   =  1+\Tr \LParen U^\dagger   \Phi_\alpha(\rho)   U \, \Delta \rho_\alpha \, \rho \RParen 
\end{align}
(App.~\ref{surpftapp}). The second term---the correction---arises from $\rho$'s coherences relative to the charges' product eigenbases. If $\rho =\rho^\as_\GGE \otimes \rho^\bs_\GGE$, as throughout much of thermodynamics, then $\rho$ can have such coherences only if charges fail to commute. 
Our correction resembles that in~\cite{Levy2020}
but arises from distinct physics: noncommuting charges, rather than initial correlations.

\subsection{Trajectory stochastic entropy production}
\label{traj}

$\straj$ evokes how entropy accompanies irreversibility (Sec.~\ref{sec_traj_def}). $\straj$ can assume nonreal values, signaling contextuality in a noncommuting-current experiment (Sec.~\ref{sec_nonclass}). Despite the unusualness of nonreal entropy production, $\straj$ satisfies two sanity checks: $\expval{\straj}$ has a sensible physical interpretation (Sec.~\ref{sec_traj_avg}), and $\straj$ obeys a correction-free fluctuation theorem (Sec.~\ref{traj_FT}).
Complex-valued entropy production appeared also in Ref. \cite{kwon2019}.

\subsubsection{Trajectory SEP formula}
\label{sec_traj_def}

$\straj$ generalizes the conventional trajectory SEP formula, which we review now. Recall the classical experiment in Sec.~\ref{sec_intro}: Classical systems $\as$ and $\bs$ begin in grand canonical ensembles, then exchange energy and particles. In each trial, $\as \bs$ undergoes the trajectory
$( E^{\as}_i, N^{\as}_i, E^{\bs}_i, N^{\bs}_i )$ $\mapsto$
$( E^{\as}_f, N^{\as}_f, E^{\bs}_f, N^{\bs}_f )$ with some joint probability. Let us add a subscript F to that probability's notation:
$p_{\rm F} ( E^{\as}_i, N^{\as}_i, E^{\bs}_i, N^{\bs}_i \, ; \, E^{\as}_f, N^{\as}_f, E^{\bs}_f, N^{\bs}_f )$.
Imagine preparing the grand canonical ensembles, then implementing the time-reversed dynamics.
One observes the reverse trajectory, 
$( E^{\as}_f, N^{\as}_f, E^{\bs}_f, N^{\bs}_f )$ $\mapsto$
$( E^{\as}_i, N^{\as}_i, E^{\bs}_i, N^{\bs}_i )$,
with a probability
$p_{\rm R}( E^{\as}_f, N^{\as}_f, E^{\bs}_f, N^{\bs}_f \, ; \, 
E^{\as}_i, N^{\as}_i, E^{\bs}_i, N^{\bs}_i )$.
The probabilities' log-ratio forms the conventional trajectory SEP formula~\cite{crooks_nonequilibrium_1998, evans_probability_1993, Gallavotti_95_Dynamical, Jarzynski2004}: $\log(p_{\rm F} / p_{\rm R})$.
In an illustrative forward trajectory, heat and particles flow from a hotter, higher-chemical-potential $\as$ to a colder, lower-chemical-potential $\bs$. This forward trajectory is likelier than its reverse: $p_{\rm F} > p_{\rm R}$. Hence $\log(p_{\rm F} / p_{\rm R}) > 0$, as expected from the second law of thermodynamics.

Let us extend this formula to quasiprobabilities.
Section~\ref{KD} established a forward protocol suitable for potentially noncommuting $Q_\alpha$'s. That section attributed to the forward trajectory
$(i_1, i_2, \ldots, i_c)$ $\mapsto$ 
$(f_c, f_{c-1}, \ldots, f_1)$
the KDQ $\tilde{p}_{\rm F} 
(i_1, i_2, \ldots, i_c \, ; \,
 f_c, f_{c-1}, \ldots, f_1)$
[Eq.~\eqref{KDdef}].
The reverse protocol features $U^\dag$, rather than $U$, in step~\ref{step_U}, with a reversed list of measurement outcomes. The reverse trajectory $(f_1, f_2, \ldots, f_c)$ $\mapsto$ 
$(i_c, i_{c-1}, \ldots, i_1)$ corresponds to the quasiprobability
$\tilde{p}_{\rm R}
 (f_1, f_2, \ldots, f_c \, ; \, 
 i_c, i_{c-1}, \ldots, i_1)
\coloneqq \Tr( [ \Pi_{1,f_1} \ldots \Pi_{c, f_c} ] U [ \Pi_{c, i_c} \ldots \Pi_{1, i_1} ] U^\dagger \rho)$.
This definition captures the notion of time reversal, we argue in App.~\ref{app_traj_rev_def}, while enabling $\straj$ to agree with $\schrg$ and $\ssurp$ in the commuting case (App.~\ref{app_traj_reduce}).
Both quasiprobabilities feature in the trajectory SEP: 
\begin{align}
   \label{eq:straj-long}
   & \straj \coloneqq
   \log \left( \frac{ 
   \tilde{p}_\text{F} 
   (i_1, i_2, \ldots, i_c \, ; \,
    f_c, f_{c-1}, \ldots, f_1) }{ 
   \tilde{p}_\text{R}
   (f_1, f_2, \ldots, f_c \, ; \, 
    i_c, i_{c-1}, \ldots, i_1) } \right) \\
   & = \log \left( \frac{\Tr \left( U^\dagger 
   \left[ \Pi_{1,f_1} \ldots \Pi_{c, f_c} \right] U 
   \left[ \Pi_{c,i_c} \ldots \Pi_{1, i_1} \right] \rho \right)}{
   \Tr \left( 
       \left[ \Pi_{1,f_1} \ldots \Pi_{c, f_c} \right] U 
   \left[ \Pi_{c, i_c} \ldots \Pi_{1, i_1} \right] U^\dagger \rho \right)} \right) 
   \nonumber \\
   &=\log \left( \frac{
   \bra{i_1} \rho U^\dagger \ket{f_1}}{
   \bra{i_1} U^\dagger \rho \ket{f_1} } \right) .
   \label{eq:straj_single_charge}
\end{align}
The final equality follows from the nondegeneracy of the local charges $Q_\alpha^{\tt X}$: The projectors $\Pi_{\alpha, k} = \ketbra{\alpha_k}{\alpha_k}$, so factors cancel between the numerator and denominator.
The $\bra{i_1}$'s and $\ket{f_1}$'s distinguish $Q_1$ from the other charges. However,~\eqref{eq:straj_single_charge} holds for every possible labeling of the $Q_\alpha$'s (holds under the labeling of any charge as 1) and for every ordering of the projectors in~\eqref{KDdef}. Moreover, App.~\ref{app_symmetrized} introduces a variation on Eq.~\eqref{eq:straj-long}---a definition that contains an average over all measurement orderings, removing the dependence on the ordering.

\subsubsection{Nonreal trajectory SEP witnesses nonclassicality} 
\label{sec_nonclass}

Nonreal $\straj$ values signal nonclassicality in a noncommuting-current experiment. To prove this result, we review weak values (conditioned expectation values) and contextuality (provable nonclassicality). 
We then express $\straj$ in terms of weak values. Finally, we prove that nonreal $\straj$ values herald contextuality in an instance of the forward or reverse protocol.

\begin{figure}[t]
\centering
\includegraphics[width=0.4\textwidth]{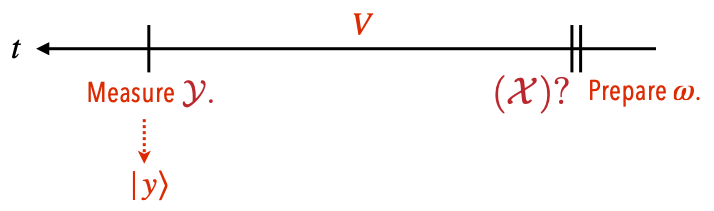}
\caption{\caphead{Weak value:} The weak value serves as a ``conditioned expectation value'' in the protocol depicted. Time runs from right to left, as reading a weak value from right to left translates into the procedure depicted: Prepare $\omega$, evolve the state under $V$, measure $\mathcal{Y}$, and postselect on outcome $y$. Which value is retrodictively most reasonably attributable to $\mathcal{X}$ immediately after the preparation procedure? Arguably the weak value~\cite{johansen_nonclassical_2004, hall_exact_2001, hall_prior_2004, dressel_weak_2015}.}
\label{fig:weak_measurement}
\end{figure}

Figure~\ref{fig:weak_measurement} motivates the weak value's form~\cite{Aharonov1988, Dressel2014}.
Consider preparing a quantum system in a state 
$\dummy$ at a time $t=0$.
The system evolves under a unitary $V$. An observable $\mathcal{Y}$ is then measured, yielding an outcome $y$. Denote by $\mathcal{X}$ an observable that neither commutes with $\dummy$ nor shares the $\mathcal{Y}$ eigenstate $\ket{y}$. Which value is retrodictively most reasonably attributable to $\mathcal{X}$ immediately after the state preparation?
Arguably the \emph{weak value}~\cite{johansen_nonclassical_2004, hall_exact_2001, hall_prior_2004, dressel_weak_2015}
$\Tr(\Pi'_y \mathcal{X} \dummy)/\Tr(\Pi_y' \omega)$,
wherein 
$\Pi'_y \coloneqq V^\dagger | y\rangle\langle y| V$.
Weak values can be \emph{anomalous}, lying outside the spectrum of $\mathcal{X}$. Anomalous weak values actuate metrological advantages~\cite{Pang2014, Jordan2014, Harris2017, Xu2020, Arvidsson-Shukur2020,DRMAS_22_Quantum} and signal contextuality~\cite{Tollaksen2007, Pusey2014, Kunjwal2019}.

\emph{Contextuality} is a strong form of nonclassicality~\cite{Kochen1968, Spekkens2005}.
One can model quantum systems as being in unknown microstates akin to classical statistical-mechanical microstates. One might expect to model operationally indistinguishable procedures identically. No such model reproduces all quantum-theory predictions, however. This impossibility is quantum theory's contextuality, which underlies quantum-computational speedups, for example~\cite{Howard2014}.
Anomalous weak values signal contextuality in the process 
\emph{prepare $\dummy$, measure $\mathcal{X}$ weakly, evolve with $V$, measure $\mathcal{Y}$ strongly, and postselect on $y$}~\cite{Tollaksen2007, Pusey2014, Kunjwal2019}.

Having introduced the relevant background, we prove that $\straj$ depends on weak values and signals contextuality. 
Define the evolved projectors
$\Pi_{1, i_1}' \coloneqq U \Pi_{1, i_1} U^\dagger$ and $\Pi_{1, f_1}'' \coloneqq U^\dagger \Pi_{1, f_1} U$.
Define also the weak values  
\begin{align}
   & {}_{f_1}\langle \Pi_{1,i_1} \rangle_{\rho}
   \coloneqq \Tr(\Pi_{1, f_1}'' \Pi_{1, i_1} \rho) / \Tr(\Pi_{1, f_1}'' \rho)
   \quad \text{and} \\ &
{}_{i_1}\langle \Pi_{1, f_1} 
   \rangle_{\rho} 
   \coloneqq \Tr(\Pi_{1, i_1}' \Pi_{1,f_1} \rho)/\Tr(\Pi_{1,i_1}' \rho).
\end{align}
Each is a complex number with a phase
$\phi$:
${}_{f_1}\langle \Pi_{1,i_1} \rangle_{\rho}
= | {}_{f_1}\langle \Pi_{1,i_1} \rangle_{\rho} |
e^{i \phi_{\rm F} }$, and
${}_{i_1}\langle \Pi_{1, f_1} 
   \rangle_{\rho}
= | {}_{i_1}\langle \Pi_{1, f_1} 
   \rangle_{\rho} |
e^{-i \phi_{\rm R} }$.
We suppress the phases' indices for conciseness. 
Equation~\eqref{eq:straj_single_charge} becomes
\begin{align}\label{eq:straj_cartesian}
   \straj 
      &=\log \left(  \big\lvert 
   {}_{f_1}\langle \Pi_{1,i_1} \rangle_{\rho} \big\rvert /
  \big\lvert 
  {}_{i_1}\langle \Pi_{1, f_1} 
   \rangle_{\rho}  \big\rvert   \right)
   + i(\phi_\text{F} - \phi_\text{R})
   \nonumber \\ & \quad \;
  + \log \LParen \Tr ( \Pi_{1, f_1}'' \rho ) /
            \Tr (\Pi_{1,i_1}' \rho ) \RParen .
\end{align}
The final log is of mere probabilities.
The phase difference is defined modulo $2 \pi$.
For convenience, we assume the complex logarithm's branch cut lies along the negative real axis.\footnote{
This choice renders $\straj$ well-defined in the commuting case.}
Appendix~\ref{app:traj_avg_pure_states} explains how to choose the complex logarithm's value if $\rho$ is pure and describes subtleties concerning mixed $\rho$'s.

If ${\rm Im}(\straj) \neq 0$, we call $\straj$ \emph{anomalous}, and at least one weak value is anomalous. Hence at least one of two protocols is contextual:
\begin{enumerate}
   \item \emph{Forward compressed protocol:} 
   Prepare $\rho$. Measure $\Pi_{1, i_1}$ weakly.\footnote{
   One can use the weak-measurement technique used in the forward protocol~\cite[footnote~9]{NYH_18_Quasiprobability}.}
   Evolve under $U$. Measure the $Q_1$ product basis strongly. Postselect on $f_1$.
   
   \item \emph{Reverse compressed protocol:} 
    Prepare $\rho$. Measure $\Pi_{1, f_1}$ 
    weakly. Evolve under $U^\dagger$. Measure the $Q_1$ product basis strongly. Postselect on $i_1$.
\end{enumerate}
The forward compressed protocol is a simplification of the forward protocol in Sec.~\ref{KD}: Only measurements pertaining to charge $Q_1$ are performed. Analogous statements concern the reverse compressed protocol. 
Hence $\straj$ joins a sparse set of thermodynamic quantities known to signal contextuality~\cite{lostaglio_certifying_2020, Levy2020}.

$\straj$'s signaling of contextuality exhibits irreversibility---fittingly for an entropic phenomenon---algebraically and geometrically. First, for $\straj$ to signal contextuality---to become nonreal---nonreality of a weak value does not suffice. Rather, the weak values' phases must fail to cancel: 
$\phi_{\rm F} \neq \phi_{\rm R}$.\footnote{
A compressed protocol can be contextual without $\straj$'s becoming nonreal also if a weak value is real but $\not\in [0, 1]$.}
This failure mirrors how, conventionally, entropy is produced only when $p_{\rm F} \neq p_{\rm R}$.

Second, suppose that $\rho$ is pure.
$\phi_{\rm F} - \phi_{\rm R}$
equals the geometric phase imprinted on a state manipulated as follows: $\rho$ is prepared, the forward compressed protocol is performed, the state is reset to $\rho$, the reverse compressed protocol is performed, and the state is reset to $\rho$~\cite{Pancharatnam1956, Samuel_General_1988}. All measurements here are performed in the strong limit. The state acquires the phase $e^{i \phi_{\rm F} }$ during the forward step and acquires $e^{-i \phi_{\rm R} }$
during the reverse. Only if the forward and reverse steps fail to cancel does the geometric phase $\neq 1$---does $\straj$ signal contextuality. 
Hence $\straj$ heralds contextuality in the presence of irreversibility, appropriately for a thermodynamic quantity.

\subsubsection{Average of trajectory SEP}
\label{sec_traj_avg}

Formally, averaging $\straj$ [Eq.~\eqref{eq:straj-long}] with respect to $\tilde{p}_{\rm F}$ yields a quasiprobabilistic relative entropy of Sec.~\ref{entr}:\footnote{
If $\rho$ is mixed, the relative entropy's value can depend on one's branch-cut convention.}
\begin{align}
   \expval{\straj}
   = D \LParen & \tilde{p}_\text{F} 
   (i_1, i_2, \ldots, i_c \, ; \,
    f_c, f_{c-1}, \ldots, f_1) 
   \\ & \nonumber
   || \, \tilde{p}_\text{R}
   (f_1, f_2, \ldots, f_c \, ; \, 
    i_c, i_{c-1}, \ldots, i_1) \RParen .
\end{align}
The average can be negative and even nonreal. Yet $\expval{\straj}$ has a particularly crisp physical interpretation when $\rho$ is a pure state that retains GGE marginals~\eqref{eq_GGE}.\footnote{
Throughout much of the paper, we assume that $\rho$ equals the tensor product~\eqref{initstate}. Here, $\rho$ differs to enhance the example's clarity.}

A pure $\rho$ is not as restricted as it may seem: Thermodynamics often features pure global states whose reduced states are thermal~\cite{Lloyd_88_Black,Goldstein_06_Canonical,Popescu_06_Entanglement,Partovi2008,Jennings2010}.
The average becomes
\begin{align}
   \label{avgt}
   \expval{\straj} 
   & = \tfrac{1}{2} \big[ 
   D\boldsymbol{(}\rho || \Phi_1( U^\dagger \rho U)\boldsymbol{)} 
   + D\boldsymbol{(}\rho||  U^\dagger \Phi_1(\rho) U\boldsymbol{)} 
   \nonumber \\ & \qquad \; \:
   - D\boldsymbol{(}\rho || \Phi_1(\rho)\boldsymbol{)}
   - D\boldsymbol{(}\rho_{\rm f} || \Phi_1(\rho_{\rm f})\boldsymbol{)} \big]  
   \nonumber \\ & \quad \,
   + i \langle \phi_\text{F} - \phi_\text{R}\rangle 
\end{align}
(App.~\ref{app:traj_avg_pure_states}).
Each $\phi$ implicitly depends on indices $i_1$ and $f_1$.
$\expval{\straj}$'s real and imaginary components have physical significances that we elucidate now.

In App.~\ref{app:traj_avg_pure_states}, we prove that $\langle \phi_\text{F} - \phi_\text{R}\rangle $ is real.\footnote{
This fact is not obvious: although the $\phi$'s are real, they are averaged with respect to $\tilde{p}_{\rm F}$, which can be nonreal.} Hence, if $\Im(\expval{\straj}) \neq 0$, at least one $\phi_{\rm F}$ [associated with one tuple $(i_1, f_1)$] or one $\phi_{\rm R}$ is nonzero. Therefore, at least one weak value---at least one 
${}_{f_1}\langle \Pi_{1, i_1}\rangle_\rho$ or ${}_{i_1}\langle \Pi_{1, f_1}\rangle_\rho$---is anomalous. At least one instance of the forward or reverse compressed protocol is therefore contextual. In conclusion, $\expval{\straj}$, beyond $\straj$, witnesses contextuality.

$\Re(\langle \straj\rangle)$ has two familiar properties. First, 
$\Re(\langle \straj\rangle) \geq 0$, suggestively of the second law of thermodynamics.
Second, $\Re(\langle \straj\rangle)$ depends on relative-entropy differences, similarly to Eq. \eqref{relentr}. 

Pairs of relative entropies have recognizable physical significances. Each negated $D$ is a relative entropy of coherence, comparing a state ($\rho$ or $\rho_{\rm f}$) to its dephased counterpart~\cite{streltsov_colloquium_2017}.
Hence states' coherences relative to 
$\{ \Pi_{1, k} \}$ reduce $\expval{\straj}$. This reduction resembles the reduction of $\expval{\ssurp}$ by initial coherence (Sec.~\ref{sec_surp_avg}). Not only initial coherences, though, but also final coherences reduce $\expval{\straj}$.

Equation~\eqref{avgt}'s first two relative entropies imprint noncommutation on $\expval{\straj}$. In each such $D$, the compared-to state (the final argument) results from a dephasing and a time-reversed evolution. The operations' ordering differs between the $D$'s. The $1/2$ in Eq.~\eqref{avgt} averages over the orderings. Hence $\expval{\straj}$ translates into sensible physics.

\subsubsection{Fluctuation theorem for trajectory SEP}
\label{traj_FT}

Passing another sanity check, $\straj$ satisfies the correction-free fluctuation theorem
\begin{align}
   \label{trajft}
   \expval{e^{-\straj}} =  1.
\end{align}
The proof follows from the KDQ's normalization.

 \section{Two-qubit example}
\label{example}

This section numerically illustrates the SEP definitions' key properties. First, we introduce the system simulated. Afterward, we analyze the calculated SEPs using generic parameter choices.

In our example system, $\as$ and $\bs$ are qubits. The charges are the Pauli operators $Q_1=\sigma_z$, $Q_2=\sigma_y$, and $Q_3=\sigma_x$.
By the Schur--Weyl duality, the most general charge-conserving unitary is a linear combination of the permutations of two objects---a linear combination of the identity and SWAP operators~\cite{Watrous2018,NYH_22_How}. The SWAP operator acts on states $\ket{\psi}_\as$ and $\ket{\phi}_\bs$ as $\text{SWAP} \ket{\psi}_\as \otimes \ket{\phi}_\bs = \ket{\phi}_\as \otimes \ket{\psi}_\bs $. We parameterize the unitary with an angle $\theta$:
\begin{equation}\label{eq:qubit_unitary}
    U_\theta = \cos (\theta) \, \id + i \sin(\theta) \ \text{SWAP} \, .
\end{equation}

\subsection{Charge SEP}

Section~\ref{sec_chrg_FT} introduced nonconserving trajectories and the following results. On the RHS of the fluctuation theorem~\eqref{noncommft} are two terms. The first, notated as $\expval{e^{-\kappa}}$ (App. \ref{app:schrg_FT}), encodes the noncommutation of the charges in the initial state [Eqs.~\eqref{eq_GGE} and~\eqref{initstate}].
The fluctuation theorem's final term, notated as 
$ \expval{e^{-\schrg}-e^{-\kappa}}_\text{noncons}$ 
(App. \ref{app:schrg_FT}), is an average over nonconserving trajectories. 

\begin{figure}[h]
\centering
\includegraphics[width=0.45\textwidth]{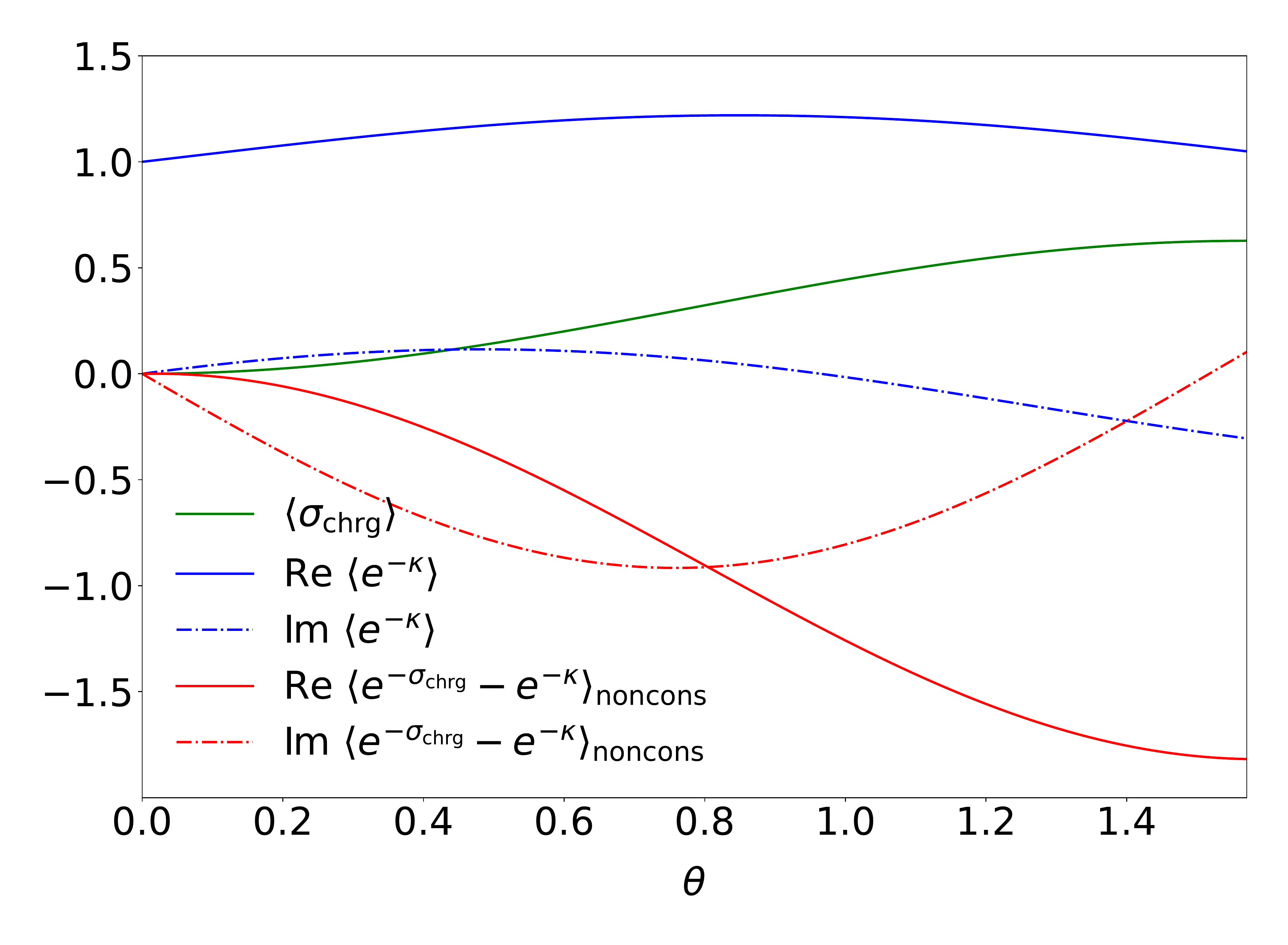}
\caption{\caphead{Components of $\schrg$ fluctuation theorem, and $\expval{\schrg}$, vs. $\theta$.} The generalized inverse temperatures are $\beta_x^\as=0.7$, $\beta_x^\bs=0.1$,  $\beta_y^\as=1$, $\beta_y^\bs=0.2$, $\beta_z^\as=0.5$, and $\beta_z^\bs=0.6$.}
\label{fig:qubit_chrg}
\end{figure}

We support these claims by showing how the two terms change as $\theta$ varies from $0$ to $\frac{\pi}{2}$ in Eq.~\eqref{eq:qubit_unitary}.
Figure~\ref{fig:qubit_chrg} shows the terms' real and imaginary parts, as well as the average entropy production $\expval{\schrg}$. The generalized inverse temperatures' values (listed in Fig.~\ref{fig:qubit_chrg}'s caption) ensure that a strong thermodynamic force pushes the $\sigma_x$ and $\sigma_y$ charges from $\tt B$ to $\tt A$, whereas a weak force pushes $\sigma_z$ oppositely. 

The unitary parameter $\theta$ increases along the $x$-axis.
As per a sanity check in Sec.~\ref{sec_3formulae}, when $\theta=0$, the fluctuation theorem's terms sum to one. The first term, $\expval{e^{-\kappa}}$, equals one; and the second term, $ \expval{e^{-\schrg}-e^{-\kappa}}_\text{noncons}$, vanishes. As $\theta$ increases, charges flow more, as evidenced by the increasing $\expval{\schrg}$. $\expval{e^{-\kappa}}$ changes little. This near-constancy reflects the origination of $\expval{e^{-\kappa}}$ in the noncommutation of the charges in the initial state, which remains constant as $\theta$ changes. In contrast, as $\theta$ grows, $ \expval{e^{-\schrg}-e^{-\kappa}}_\text{noncons}$ increases in magnitude---due to both its real and imaginary parts. This growth reflects nonconserving trajectories' growing contribution to the fluctuation theorem's RHS.

\subsection{Surprisal SEP}
The initial state $\rho$ can have coherences relative to each charge's product basis. As discussed in Sec.~\ref{sec_surp_avg}, these coherences  can reduce $\langle \ssurp \rangle$, even rendering it negative. $\ssurp$ is defined in terms of an arbitrary charge index $\alpha$, which we choose to be 1: $Q_1 = \sigma_z$.

\begin{figure}[h]
\centering
\includegraphics[width=0.45\textwidth]{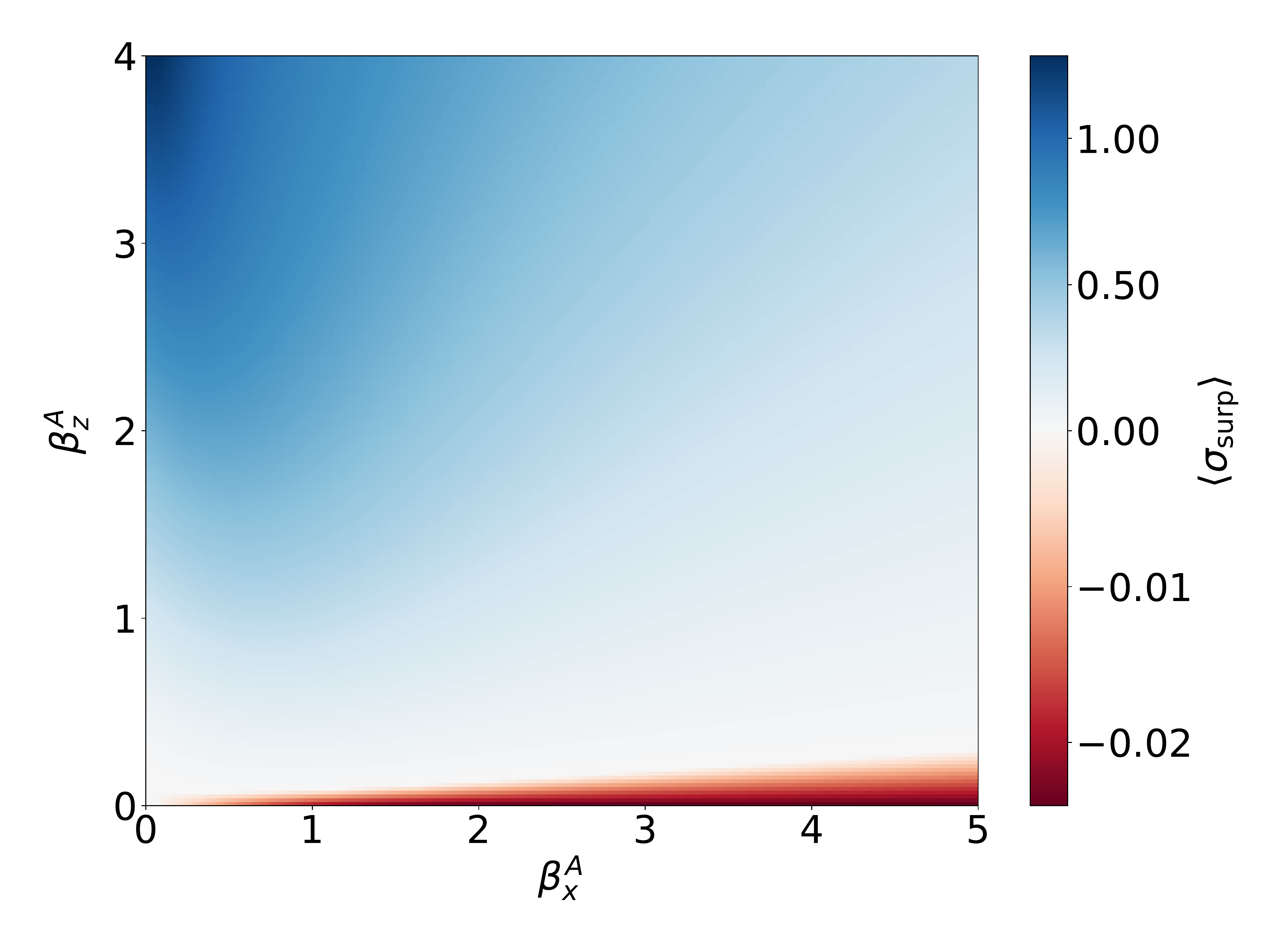}
\caption{\caphead{$\langle \sigma_{\rm surp}\rangle$ as a function of generalized inverse temperatures $\beta_x^\as$ and $\beta_z^\as$.} The unswept parameters are $\beta_y^{\tt A} = 0$, $\beta_x^{\tt B} = 0$, $\beta_y^{\tt B} = 1.6$, $\beta_z^{\tt B} = 0.1$, and $\theta = \pi/5$.}
\label{fig:qubit_surp}
\end{figure}

Figure~\ref{fig:qubit_surp} shows $\expval{\ssurp}$ as a function of $\beta_x^{\tt A}$ and $\beta_z^{\tt A}$. If $\beta_z^{\tt A} \gg \beta_x^{\tt A}$, $\rho^{\as}$ is nearly diagonal relative to the $\sigma_z$ eigenbasis. 
Hence $\expval{\ssurp}$ is positive, as evidenced in the plot's top left-hand corner. In the opposite regime ($\beta_x^{\tt A} \gg \beta_z^{\tt A}$), $\rho^{\as}$ has large coherences relative to the $\sigma_z$ eigenbasis. These coherences drive $\expval{\ssurp}$ below zero, as evidenced in the bottom right-hand corner.

\subsection{Trajectory SEP}
As shown in Sec.~\ref{sec_nonclass}, a nonreal $\straj$ implies contextuality in a noncommuting-charge experiment.
$\straj$ carries indices $i_1$ and $f_1$, by the definition~\eqref{eq:straj-long}. We numerically calculate the $\straj$ evaluated on the trajectory defined by $\ket{i_1} = \ket{0}^\as \otimes \ket{1}^\bs$ and $\ket{f_1} = \ket{0}^\as \otimes \ket{0}^\bs $.

\begin{figure}[h]
\centering
\includegraphics[width=0.45\textwidth]{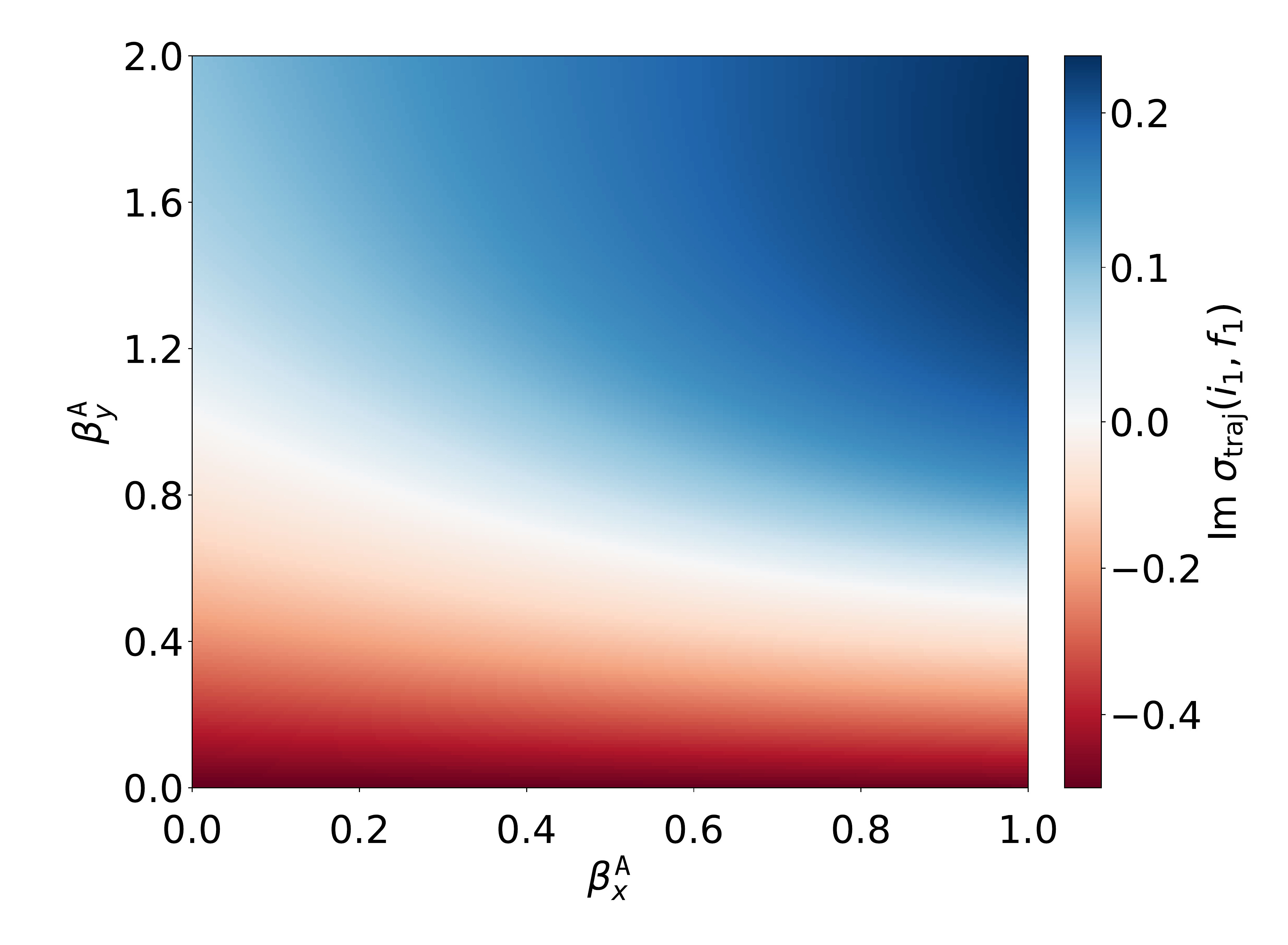}
\caption{\caphead{Imaginary part of $\sigma_{\rm traj}$, plotted against $\beta_x^{\tt A}$ and $\beta_y^{\tt A}$.} 
The unswept parameters are $\beta_z^{\tt A} = 0.01$, $\beta_x^{\tt B} = 0.01$, $\beta_y^{\tt B} = 1$, $\beta_z^{\tt B} = 0.01$, and $\theta=0.5$.}
\label{fig:qubit_traj}
\end{figure}

Figure~\ref{fig:qubit_traj} shows the imaginary part of ${\straj} (i_1,f_1)$ plotted against $\beta_x^{\tt A}$ and $\beta_y^{\tt A}$.
$\sigma_{\rm traj}$ typically has a nonzero imaginary component (of a magnitude similar to the real part's), signaling contextuality. The imaginary component's magnitude grows particularly large when $\beta_y^{\tt A} \approx 0$.
Furthermore, the imaginary component exhibits stability, changing smoothly with the swept parameters.

\section{Experimental sketch}
\label{sec_experiment}

Our results can be tested experimentally. Several pieces of evidence indicate such an experiment's feasibility. First, a trapped-ion experiment recently initiated the experimental testing of noncommuting-charge thermodynamics~\cite{Kranzl_23_Experimental}.
Second, other platforms have been argued to support such tests~\cite{nyh2020noncommuting,NYH_22_How}. Examples include superconducting qubits, neutral atoms, and possibly nuclear-magnetic-resonance systems. Third, sequential weak measurements have been realized with trapped ions~\cite{pan2020weak}, superconducting qubits~\cite{wang_observing_2022}, and photonics~\cite{thekkadath_direct_2016, piacentini_measuring_2016, suzuki_observation_2016, kim_direct_2018, chen_experimental_2019}. Fourth, we now sketch a trapped-ion experiment inspired by Sec.~\ref{example}.
We outline the setup, preparation procedure, evolution, measurement, and data processing. For concreteness, we tailor the proposal to the platform reported on in~\cite{Green2022}. 

The platform consists of ${}^{171}\text{Yb}^+$ ions in a linear Paul trap. 
Each ion encodes a qubit in two hyperfine ground states (two energy levels that result from splitting a ground space with a hyperfine interaction).
Our proposal calls for ten qubits: Two form the system of interest, two ancillas enable state preparation, and six ancillas enable weak measurements. 
The system of interest consists of a qubit ${\tt A}$ and a qubit ${\tt B}$. These qubits will exchange charges $\sigma_x$, $\sigma_y$, and $\sigma_z$. However, tracking just two noncommuting charges suffices for an initial experiment. We choose $\sigma_x$ and $\sigma_z$.

${\tt A}$ and ${\tt B}$ must be prepared in a tensor product of GGEs [Eq.~\eqref{initstate}]. One procedure involves two ancilla qubits: Prepare ${\tt A}$ and an ancilla in a thermofield-double state (a purification of a GGE), using the trapped-ion protocol in~\cite{Green2022}. Discard the ancilla qubit. Repeat these two steps with qubit ${\tt B}$ and another ancilla. Generalized inverse temperatures $\beta^{\tt X}_\alpha$ parameterize the initial state. One chooses the parameters' values as in Sec.~\ref{example}, to observe the three results listed two paragraphs below.

After the state preparation, one weakly measures $\sigma_x^{\tt A}$ and $\sigma_x^{\tt B}$, then $\sigma_z^{\tt A}$ and $\sigma_z^{\tt B}$.
One can implement a weak measurement using a qubit ancilla, using the circuit shown in Fig.~1(a) of~\cite{Dressel2018Strengthening}. Hence the initial weak measurements require four ancillas, total. 

After these weak measurements, ${\tt A}$ and ${\tt B}$ evolve under a charge-preserving unitary $U_\theta$ [Eq.~\eqref{eq:qubit_unitary}]. The trapped-ion platform under consideration offers a gate set formed from arbitrary single-qubit rotations and $XX$ gates~\cite{Debnath2016, Green2022}. The gate set's universality implies that $U_\theta$ can be implemented, if the ions retain coherence for long enough.
The two-qubit gate requires the most time---between 1 and 100s of $\mu$s~\cite{Debnath2016}. However, coherence times range from 100s of ms to 100s of seconds.
%%% NYH: Do not delete: reference for the previous two sentences: 4/26/24 message from Alaina Green in email chain "Experimental-proposal sketch in theory paper"
%
% single-qubit coherence times extend to seconds or even minutes~\cite{Kotler2014magnetic, Wang2017long}. 
The time scales are significantly separated, although gate errors will further restrict circuit depth~\cite{Linke_17_Experimental}.

After the evolution, one weakly measures $\sigma_x^\as$ and $\sigma_x^\bs$. Our abstract protocol (Sec.~\ref{KD}) ends with weak measurements of $\sigma_z^\as$ and $\sigma_z^\bs$. An experimentalist can replace these weak measurements with strong measurements, without hindering the reconstruction of $\tilde{p}_{\rm F}$ from the experimental data~\cite{NYH_17_Jarzynski}. 
Furthermore, the replacement spares us the need for two extra ancillas.

One repeats the foregoing protocol many times. From the measurement outcomes, we infer the probability distribution over the possible measurement outcomes. One reconstructs the KDQ $\tilde{p}_{\rm F}$ via the procedure in~\cite[App.~A]{NYH_17_Jarzynski}. 
Analogously, one infers $\tilde{p}_{\rm R}$ from another batch of trials, guided by the reverse protocol of Sec.~\ref{sec_traj_def}. Since our theoretical results are deductive, the proposed experiment essentially checks quantum theory's accuracy. However, observing three phenomena would highlight quantum features exhibited by the SEPs when charges fail to commute: (i) Observe a violation of detailed charge conservation. (ii) Observe a negative $\expval{\ssurp}$. (iii) Observe an imaginary component of $\straj$.

\section{Outlook}\label{conc}

Noncommuting charges challenge common expectations about entropy production.
Three common SEP formulae, though equal when charges commute, separate when charges do not. 
The formulae also offer different physical insights into how noncommuting charges impact entropy production.
First, the noncommutation enables stochastic trajectories to violate charge conservation individually. We introduce these nonconserving trajectories, possible only if charges fail to commute, as quantum phenomena in stochastic thermodynamics.
The violations of detailed charge conservation underlie commutator-dependent corrections to a fluctuation theorem. 
Second, initial coherences relative to charges' eigenbases can render $\expval{\ssurp}$ negative. A common (two-thermal-reservoir) setup can entail such coherences only if the charges fail to commute. Hence, charges' noncommutation source a resource that can, in a sense, effectively reverse time's arrow.  
Third, nonreality of $\straj$ signals contextuality---provable nonclassicality---in a noncommuting-current experiment. 
Such thermodynamic signatures of contextuality---a stringent criterion for nonclassicality---are rare. 
These results hold arbitrarily far from equilibrium. 
In addition to proving these results deductively, we illustrated them numerically and sketched a trapped-ion test.

Our work introduces noncommuting charges into stochastic thermodynamics~\cite{Strasberg_22_Quantum}. The field's results now merit reevaluation, in case charges' noncommutation alters them. 
For example, thermodynamic uncertainty relations bound a current's relative variance with entropy production~\cite{Barato15Thermodynamic, Gingrich2016Dissipation, timpanaro2019thermodynamic, Horowitz2020Thermodynamic}. Lowering entropy production, noncommuting charges may increase the relative variance, increasing currents' unpredictability. Our work therefore motivates the derivation of thermodynamic uncertainty relations that highlight exchanges of noncommuting charges, using KDQs. Such relations may follow as extensions of~\cite{timpanaro2019thermodynamic}.

As another avenue opened by our work within stochastic thermodynamics, nonconserving trajectories provide a new tool for unearthing genuinely quantum effects. Consider any charge exchange modeled with a KDQ. The KDQ decomposes into a conserving and nonconserving part. The former is the only piece that survives when the process has a classical (probabilistic) description. The average of every stochastic physical variable (analogous to entropy production), with respect to the KDQ, decomposes likewise. Hence the contribution of charge noncommutation to the average can be delineated clearly. For example, this technique could be applied to decompose the averaged stochastic work performed by a monitored quantum engine operating between noncommuting-charge reservoirs.

Quantum engines offer another possible application~\cite{Kosuke2018}: 
The more entropy an engine produces, the lesser the engine's efficiency. We showed that charges' noncommutation can lower entropy production even on average: Coherences relative to charges' eigenbases serve as a resource for reducing average entropy production (Sec.~\ref{surp}). Hence engines might leverage such coherences. Such an engine could exchange not only heat, but also noncommuting charges, with reservoirs. One could extend the engine, proposed in~\cite{Wright_18_Quantum}, that leverages a spin reservoir in place of a heat reservoir.

To put the charges on even more-equal footing, one could average over them in the KDQ's and SEPs' definitions (App.~\ref{app_symmetrized}). 
Finally, calculating a closed form for the average trajectory SEP, for mixed global states, remains an open problem.

Experimental opportunities complement the theoretical ones. 
In Sec.~\ref{sec_experiment}, we sketched a trapped-ion experiment for testing our results. Such a test would highlight the quantum features manifested by SEPs when charges fail to commute.

\begin{acknowledgments}
The authors thank Alaina Green, Norbert Linke, Nhung H. Nguyen, and Yingyue Zhu for discussions about trapped ions.  
This work received support from 
the John Templeton Foundation (award no. 62422).
The opinions expressed in this publication are those of the authors and do not necessarily reflect the views of the John Templeton Foundation or UMD.
T.U. acknowledges the support of the Joint Center for Quantum Information and Computer Science through the Lanczos Fellowship, as well as the Natural Sciences and Engineering Research Council of Canada (NSERC), through the Doctoral Postgraduate Scholarship.
\end{acknowledgments}

\onecolumngrid

\begin{appendices}

\renewcommand{\thesection}{\Alph{section}}
\renewcommand{\thesubsection}{\Alph{section} \arabic{subsection}}
\renewcommand{\thesubsubsection}{\Alph{section} \arabic{subsection} \roman{subsubsection}}

\makeatletter\@addtoreset{equation}{section}
\def\theequation{\thesection\arabic{equation}}

\section{Measuring the Kirkwood--Dirac quasiprobability via the forward protocol}

\label{app_Motivate_KDQ}

Here, we show how the forward protocol (Sec.~\ref{bgsetup}) leads to the KDQ $\tilde{p}_{\rm F}$ [Eq.~\eqref{KDdef}]. One can infer $\tilde{p}_{\rm F}$ from experiments by performing the forward protocol and simpler protocols.\footnote{
% < f >
This appendix articulates the forward protocol's physical significance. If one wishes, instead, merely to infer $\tilde{p}_{\rm F}$, then one can use any KDQ-measurement protocol, which need not involve weak measurements~\cite{NYH_18_Quasiprobability,Mohseninia_19_Optimizing,Hernandez_22_Projective}.}
% < /f >
The proof hinges on weak measurements.

We briefly review a model for weak measurements~\cite{Aharonov1988,dressel_weak_2015,White_16_Preserving,NYH_17_Jarzynski,NYH_18_Quasiprobability}. During any measurement, one prepares a detector, couples the detector to the system $\mathcal{S}$ of interest, and projects the detector (measures it strongly). The outcome implies information about $\mathcal{S}$. How much information depends on the coupling's strength and duration. The entire measurement process evolves $\mathcal{S}$'s state under Kraus operators. 

\emph{Kraus operators} $K_j$ model general quantum operations~\cite{Watrous2018}. They satisfy the normalization condition 
$\sum_j K_j^\dag K_j = \id$.
Modeling a measurement that yields outcome $j$, the operators evolve a measured state $\dummy$ as
$\dummy \mapsto K_j \dummy K_j^\dag / \Tr( K_j \dummy K_j^\dag)$.
For example, the forward protocol's first weak measurement effects a Kraus operator 
\begin{align}
   \label{eq_Kraus}
   K_{1,i_1} 
   \approx ({\rm const.}) \, \id
   + g_{1,i_1}  
   \Pi_{1,i_1} \, .
\end{align}
The dimensionless coupling strength $g_{1,i_1} \in \mathbb{C}$ has a magnitude much smaller than 1.\footnote{
In the strong-measurement limit, $|g_{1,i_1}|$ is not much smaller than 1, and the approximation~\eqref{eq_Kraus} is inaccurate. To model this limit, one should return to the system-and-detector unitary from which~\eqref{eq_Kraus} is derived.}
Hence the forward protocol evolves $\rho$ to the (un-normalized) conditional state
\begin{align}
   \label{eq_meas_KD_help1}
   & \left[ K_{1, f_1} K_{2, f_2} \ldots
          K_{c, f_c} \right]
   U
   \left[ K_{c, i_c} \ldots K_{2, i_2}
          K_{1, i_1} \right]
   \rho 
   \left[ K_{c, i_c} \ldots K_{2, i_2}
          K_{1, i_1} \right]^\dag
   U^\dag
   \left[ K_{1, f_1} K_{2, f_2} \ldots
          K_{c, f_c} \right]^\dag .
\end{align}
This expression's trace equals the probability that, upon projecting all the detectors, one obtains the outcomes associated with $i_1$, $i_2$, etc.
We can substitute in for each $K$ from equations of the form~\eqref{eq_Kraus}. Then, we multiply out the factors. In the resulting sum, one term contains $2c$ projectors $\Pi$ leftward of $\rho$ and $2c$ identity operators $\id$ rightward of $\rho$. That term is the real or imaginary part of $\tilde{p}_{\rm F}$ [Eq.~\eqref{KDdef}], depending on whether the $g$'s are real or imaginary.
$\tilde{p}_{\rm F}$ is an \emph{extended KDQ}, containing $>2$ projectors~\cite{NYH_18_Quasiprobability}. However, we call $\tilde{p}_{\rm F}$ a KDQ for conciseness.

One can replace the final weak measurement with a strong measurement, for experimental convenience. The outermost Kraus operators in~\eqref{eq_meas_KD_help1}---the $K_{1, f_1}$ and $K_{1,f_1}^\dag$---will become $\Pi_{1, f_1}$'s. The trace will contain a term
$\Tr( [\Pi_{1,f_1} \Pi_{2, f_2} \ldots \Pi_{c,f_c} ]
U [\Pi_{c, i_c} \ldots \Pi_{2, i_2} \Pi_{1, i_1} ]
\rho U^\dag \Pi_{1,f_1})$.
The rightmost projector can cycle around to become the leftmost. Since
$\Pi_{1,f_1} \Pi_{1,f_1} = \Pi_{1,f_1}$,
$\tilde{p}_{\rm F}$ [Eq.~\eqref{KDdef}] again results.

\section{Degenerate charges}
\label{app_degen}

This appendix concerns generalizations of our results to degenerate charges $Q_\alpha$. 
Charge degeneracies would affect the definitions of the KDQ~\eqref{KDdef}, $\tilde{p}_{\rm F}$, and the surprisal SEP~\eqref{ssurp}, $\ssurp$. 
Each quantity is defined in terms of projectors---at least one charge's product basis, $\{ \Pi_{\alpha, k} \}$. If a $Q_\alpha$ is degenerate, at least one of its eigenprojectors $\Pi_{\alpha, j}$ will have rank $>1$. We have two choices of projectors to use in defining $\tilde{p}_{\rm F}$ and $\ssurp$: We can use degenerate projectors or pick one-dimensional projectors.
Each strategy has a drawback, although apparently not due to charges' noncommutation: The drawbacks arise even in the commuting and classical cases. Future work can elucidate degeneracies in greater detail and identify whether degeneracies can play a special role in the noncommuting case. 

According to the first strategy, we continue to use the charges' eigenprojectors, regardless of ranks.
$\tilde{p}_{\rm F}$ and $\ssurp$ would retain their definitions,~\eqref{KDdef} and~\eqref{ssurp}.
Awkwardly, the SEP formulae would no longer equal each other in the commuting case, even if $\rho$ were diagonal with respect to the charges' shared eigenbasis. 

To illustrate, we suppose that the dynamics conserve only one charge. We drop its index 1 from eigenvalues $\lambda_{1,k}$ and from eigenprojectors. According to Eq.~\eqref{schrg}, the charge SEP is $\schrg = \beta^\as \left( \lambda_{f^\as} - \lambda_{i^\as} \right) + \beta^\bs \left( \lambda_{f^\bs} - \lambda_{i^\bs} \right)$. 
In contrast, the surprisal SEP is
\begin{align}
   \ssurp 
   &= \log \left( \frac{\Tr \left( 
   \left[ \Pi_{i^\as}^\as \otimes \Pi_{i^\bs }^\bs \right] \rho \right) } {
   \Tr \left( [\Pi_{f^\as}^\as \otimes \Pi_{f^\bs }^\bs] \rho \right) } \right) \\
   &= \log \Big(
   \exp \LParen \beta^\as 
   \left[ \rank(\Pi_{f^\as} )  \lambda_{f^\as} 
          - \rank(\Pi_{i^\as})  \lambda_{i^\as} \right] \RParen \;
   \exp \LParen \beta^\bs 
   \left[ \rank(\Pi_{f^\bs} )  \lambda_{f^\bs} 
          - \rank(\Pi_{i^\bs})  \lambda_{i^\bs} \right] \RParen
   \Big) \\
   &= \beta^\as \left[ 
   \rank(\Pi_{f^\as} )  \lambda_{f^\as} 
   - \rank(\Pi_{i^\as})  \lambda_{i^\as} \right]
   + \beta^\bs \left[ \rank(\Pi_{f^\bs} )  \lambda_{f^\bs} - \rank(\Pi_{i^\bs})  \lambda_{i^\bs} \right] .
\end{align}
Due to the rank factors, $\ssurp \neq \schrg$. 

Second, we can define $\tilde{p}_{\rm F}$ and $\ssurp$ in terms of rank-one projectors only. If a $Q_\alpha$ has a degenerate eigenspace, we must choose an eigenbasis for the space. The SEP formulae will remain equal in the commuting case. However, different choices of projectors engender different correction terms in the $\schrg$ fluctuation theorem~\eqref{noncommft} and in the $\ssurp$ fluctuation theorem~\eqref{surpft}. The corrections' varying with the basis choice suggests unphysicality.

For example, let each of $\as$ and $\bs$ be a qubit. We express operators relative to an arbitrary basis: $Q_1 = \begin{pmatrix}
 1 & 0 \\
0 & 1
\end{pmatrix}$,
and $Q_2= \begin{pmatrix}
 1 & 0 \\
0 & 2
\end{pmatrix}$. 
Choosing $\Pi_{1,1} = \begin{pmatrix}
 1 & 0 \\
0 & 0
\end{pmatrix}$ and $\Pi_{1,2} = \begin{pmatrix}
 0 & 0 \\
0 & 1
\end{pmatrix}$ engenders different correction terms than 
$\Pi_{1,1} = \frac{1}{\sqrt{2}} \begin{pmatrix}
 1 & 1 \\
1 & 1
\end{pmatrix}$ and $\Pi_{1,2} = \frac{1}{\sqrt{2}} \begin{pmatrix}
 1 & -1 \\
-1 & 1
\end{pmatrix}$.

\section{Charge stochastic entropy production}\label{chrgapp}

This appendix concerns $\schrg$ (Sec.~\ref{sec_chrg_def}). 
We prove the fluctuation theorem~\eqref{noncommft} in App.~\ref{app:schrg_FT},
and explain detailed charge conservation in App.~\ref{app:detailed}.

\subsection{Proof of the charge fluctuation theorem}
\label{app:schrg_FT}

Here, we prove the $\schrg$ fluctuation theorem [Eq.~\eqref{noncommft} in Sec.~\ref{sec_chrg_FT}]:
\begin{align}\label{noncommft_app}
   & \expval{e^{-\schrg}} 
   = \Tr( U^\dagger   e^{-\Delta \beta_1 Q_1^{\as}}  \ldots e^{-\Delta \beta_c Q_c^{\as} } U e^{\Delta \beta_c Q_c^{\as}} \ldots e^{\Delta \beta_1 Q_1^{\as}} \rho ) 
   + \left(  \text{$c-1$ terms dependent on
   commutators} \right) .
\end{align}
Let us add and subtract
$\sum_{\alpha=1}^c \beta_\alpha^\bs(\lambda_{\alpha, i_\alpha^\as} - \lambda_{\alpha, f_\alpha^\as})$ to and from the right-hand side of Eq.~\eqref{schrg}: 
\begin{align}
   \schrg
   & = - \sum_{\alpha=1}^c \left[\beta^\bs_\alpha ( \lambda_{\alpha, i_\alpha^\as} + \lambda_{\alpha, i_\alpha^\bs} - \lambda_{\alpha, f_\alpha^\as} - \lambda_{\alpha, f_\alpha^\bs}) + \Delta \beta_\alpha ( \lambda_{\alpha, i_\alpha^\as} - \lambda_{\alpha, f_\alpha^\as} ) \right]\, . \label{nonconscont}
\end{align}

We first calculate the fluctuation theorem's right-hand side in the commuting case.
The first parenthesized term in Eq.~\eqref{nonconscont} vanishes when evaluated on conserving trajectories (App.~\ref{app:detailed}). 
If the charges commute, therefore,
\begin{align}
    & \langle e^{-\schrg}\rangle 
    = \expval{e^{ \sum_{\alpha=1}^c \Delta \beta_\alpha \left(\lambda_{\alpha, i_\alpha^\as} - \lambda_{\alpha, f_\alpha^\as}\right) }} \label{chrgftfirsterm}\\
\nonumber  
&= \sum_{\substack{ i_1^\as , i_1^\bs ,\ldots, i_c^\as,i_c^\bs,  \\ f_1^\as , f_1^\bs ,\ldots, f_c^\as,f_c^\bs  }} 
\Tr( U^\dagger 
\left[ \Pi_{1, f^\as_1} \otimes \Pi_{1, f^\bs_1 } \right] \ldots \left[\Pi_{c, f^\as_c } \otimes \Pi_{c, f^\bs_c } \right]  U \left[\Pi_{c, i^\as_c }\otimes \Pi_{c, i^\bs_c} \right] \ldots \left[\Pi_{1, i^\as_1}\otimes \Pi_{1, i^\bs_1} \right] \rho )  e^{\sum_{\alpha=1}^c \Delta \beta_\alpha \left(\lambda_{\alpha, i_\alpha^\as} - \lambda_{\alpha, f_\alpha^\as}\right) }\\
\nonumber  
&=\sum_{\substack{ i_1^\as , i_2^\as ,\ldots, i_c^\as,  \\ f_1^\as , f_2^\as ,\ldots, f_c^\as  }} 
\Tr( U^\dagger 
\left[\Pi_{1, f^\as_1} \otimes \mathbbm{1}^\bs \right] \ldots 
\left[\Pi_{c, f^\as_c } \otimes \mathbbm{1}^\bs \right]  U 
\left[\Pi_{c, i^\as_c }\otimes \mathbbm{1}^\bs \right] \ldots 
\left[\Pi_{1, i^\as_1}\otimes \mathbbm{1}^\bs \right] \rho )  e^{\sum_{\alpha=1}^c \Delta \beta_\alpha \left(\lambda_{\alpha, i_\alpha^\as} - \lambda_{\alpha, f_\alpha^\as}\right) }\\
\nonumber 
&= 
\sum_{\substack{ i_1^\as , i_2^\as ,\ldots, i_c^\as,  \\ f_1^\as , f_2^\as ,\ldots, f_c^\as  }} 
\Tr( U^\dagger 
\left[ e^{-\Delta \beta_1 \lambda_{1, f_1^\as}} \Pi_{1, f^\as_1} \otimes \mathbbm{1}^\bs \right] \ldots \left[ e^{-\Delta \beta_c \lambda_{c, f_c^\as}} \Pi_{c, f^\as_c } \otimes \mathbbm{1}^\bs \right]  U \left[ e^{\Delta \beta_c \lambda_{c, i_c^\as}} \Pi_{c, i^\as_c }\otimes \mathbbm{1}^\bs \right] \ldots \left[ e^{\Delta \beta_1 \lambda_{1, i_1^\as}}\Pi_{1, i^\as_1}\otimes \mathbbm{1}^\bs \right] \rho )  \\
&=  \Tr( U^\dagger  \left\{ 
\left[ e^{-\Delta \beta_1 Q_1^\as} \ldots  e^{-\Delta \beta_c Q_c^\as } \right] \otimes \mathbbm{1}^\bs\right\} \, U \, \left\{ 
\left[ e^{\Delta \beta_c Q_c^\as} \ldots e^{\Delta \beta_1 Q_1^\as} \right] \otimes \mathbbm{1}^\bs \right\} \, \rho ). \label{eq:charge-FT-commuting}  
\end{align}
This expression is the fluctuation theorem's RHS in the commuting case (and equals 1 when $\rho$ is the usual tensor product of thermal states). 

We now compute $\langle e^{-\schrg}\rangle$ in the noncommuting case. To identify a correction, we separate out a term of the form~\eqref{eq:charge-FT-commuting}. 
All other terms will be commutator-dependent corrections.
First, we insert the definitions of $\schrg$ [Eq.~\eqref{schrg}] and $\tilde{p}_\text{F}$ [Eq.~\eqref{KDdef}] into the fluctuation theorem's left-hand side (LHS).
For conciseness, we suppress the indices $(i_1^\as , i_1^\bs ,\ldots, i_c^\as,i_c^\bs \ ;  \: f_1^\as , f_1^\bs ,\ldots, f_c^\as,f_c^\bs)$ in the sum:
\begin{align}
   & \expval{e^{-\schrg}}
   = \sum \Tr( U^\dagger  \Pi_{1, f_1} \ldots \Pi_{c, f_c }    U \Pi_{c, i_c } \ldots \Pi_{1, i_1 } \rho ) e^{\sum_{\alpha=1}^c \beta_\alpha^\as \lambda_{\alpha, i_\alpha^\as} + \beta_\alpha^\bs \lambda_{\alpha, i_\alpha^\bs} - \beta_\alpha^\as \lambda_{\alpha, f_\alpha^\as} - \beta_\alpha^\bs \lambda_{\alpha, f_\alpha^\bs} }\\
   &=\sum \Tr \left( U^\dagger  
   \left[ e^{- \beta_1^\as \lambda_{1, f_1^\as} - \beta_1^\bs \lambda_{1, f_1^\bs} }
   \Pi_{1, f_1 } \right]
   \ldots 
   \left[ e^{- \beta_c^\as \lambda_{c, f_c^\as} - \beta_c^\bs \lambda_{c, f_c^\bs} } 
   \Pi_{c, f_c } \right]  
   U 
   \left[ e^{\beta_c^\as \lambda_{c, i_c^\as}+ \beta_c^\bs \lambda_{c, i_c^\bs}  } 
   \Pi_{c, i_c } \right] 
   \ldots 
   \left[ e^{ \beta_1^\as \lambda_{1, i_1^\as}+ \beta_1^\bs \lambda_{1, i_1^\bs}  }
   \Pi_{1, i_1 } \right] 
   \rho \right)  \nonumber \\
   \label{eq_chrg_FT_help1}
   &=  \Tr( U^\dagger e^{- \beta_1^\as Q_1^\as - \beta_1^\bs Q_1^\bs} \ldots e^{- \beta_c^\as Q_c^\as - \beta_c^\bs Q_c^\bs} U e^{\beta_c^\as Q_c^\as+ \beta_c^\bs Q_c^\bs}  \ldots e^{ \beta_1^\as Q_1^\as+ \beta_1^\bs Q_1^\bs} \rho ).
\end{align}

We now massage this expression to bring out the commutator-dependent corrections.
As always, $\id$ operators are implicitly tensored on wherever necessary.
We replace the expressions  
$- \beta_\alpha^\as Q_\alpha^\as - \beta_\alpha^\bs Q_\alpha^\bs$ with
$-\Delta \beta_\alpha^\as Q_\alpha^\as - \beta_\alpha^\bs Q_\alpha^\text{tot} $
in~\eqref{eq_chrg_FT_help1}:
\begin{align}
\expval{e^{-\schrg}}
   &= \Tr( U^\dagger  
   \left[ e^{- \Delta \beta_1 Q_1^\as - \beta_1^\bs Q_1^\text{tot}} \ldots e^{- \Delta \beta_c Q_c^\as - \beta_c^\bs Q_c^\text{tot}} \right] U 
   \left[ e^{\Delta \beta_c Q_c^\as+ \beta_c^\bs Q_c^\text{tot}}  \ldots e^{ \Delta \beta_1 Q_1^\as+ \beta_1^\bs Q_1^\text{tot}} \right] 
   \rho ).
\end{align}
Now, we commute each $e^{- \beta_{\alpha}^\bs Q_{\alpha}^\text{tot}}$ from the left of $U$ to the right,
until the exponential cancels with its counterpart, $e^{ \beta_{\alpha}^\bs Q_{\alpha}^\text{tot}}$. 
Here are the first few 
manipulations, starting with the $e^{- \beta_{c-1}^\bs Q_{c-1}^\text{tot}}$ in the leftmost half of the trace's argument:
\begin{align}
& \expval{ e^{-\schrg} }
= \Tr( U^\dagger e^{- \Delta \beta_1 Q_1^\as - \beta_1^\bs Q_1^\text{tot}} \ldots
e^{- \Delta \beta_{c-1} Q_{c-1}^\as - \beta_{c-1}^\bs Q_{c-1}^\text{tot}}  
e^{- \Delta \beta_c Q_c^\as} U e^{\Delta \beta_c Q_c^\as} e^{ \Delta \beta_{c-1} Q_{c-1}^\as + \beta_{c-1}^\bs Q_{c-1}^\text{tot}} \ldots e^{ \Delta \beta_1 Q_1^\as+ \beta_1^\bs Q_1^\text{tot}} \rho ) \nonumber \\ 
& = \Tr( U^\dagger e^{- \Delta \beta_1 Q_1^\as - \beta_1^\bs Q_1^\text{tot}} \ldots
e^{- \Delta \beta_{c-1} Q_{c-1}^\as }  
e^{- \Delta \beta_c Q_c^\as} 
U
e^{\Delta \beta_c Q_c^\as} e^{ \Delta \beta_{c-1} Q_{c-1}^\as } \ldots e^{ \Delta \beta_1 Q_1^\as+ \beta_1^\bs Q_1^\text{tot}} \rho ) \\ 
&\quad + \hbox{Tr}\Big( U^\dagger e^{- \Delta \beta_1 Q_1^\as - \beta_1^\bs Q_1^\text{tot}}\ldots
e^{- \Delta \beta_{c-1} Q_{c-1}^\as} \left[e^{- \beta_{c-1}^\bs Q_{c-1}^\text{tot}},  
e^{- \Delta \beta_c Q_c^\as} 
U
e^{\Delta \beta_c Q_c^\as}\right] 
e^{ \Delta \beta_{c-1} Q_{c-1}^\as + \beta_{c-1}^\bs Q_{c-1}^\text{tot}} \ldots e^{ \Delta \beta_1 Q_1^\as+ \beta_1^\bs Q_1^\text{tot}} \rho \Big). \nonumber
\end{align}
To arrive at the second line, we canceled the exponentials that contained $Q_c^\text{tot}$. In the third line, canceling the exponentials involving $Q_{c-1}^\text{tot}$, we had to swap $e^{- \beta_{c-1}^\bs Q_{c-1}^\text{tot}}$
and $e^{- \Delta \beta_c Q_c^\as} 
U
e^{\Delta \beta_c Q_c^\as}$.
This swap induced the commutator-containing summand.
Continuing in this fashion---bringing all the $Q_\alpha^\text{tot}$-dependent exponentials together and inducing a commutator each time---we arrive at the desired form:
\begin{align}
\label{eq_chrg_ft_help1}
\expval{e^{\sum_{\alpha=1}^c \beta_\alpha^\as \lambda_{\alpha, i_\alpha^\as} + \beta_\alpha^\bs \lambda_{\alpha, i_\alpha^\bs} - \beta_\alpha^\as \lambda_{\alpha, f_\alpha^\as} - \beta_\alpha^\bs \lambda_{\alpha, f_\alpha^\bs} }} 
& =   \Tr( U^\dagger e^{-\Delta \beta_1 Q_1^{\as} } \ldots e^{-\Delta \beta_c Q_c^{\as}}  U e^{\Delta \beta_c Q_c^{\as}} \ldots e^{\Delta \beta_1 Q_1^{\as}} \rho )  \\ & \quad \;
+ ( \text{$c-1$ commutator-dependent terms}). 
\end{align}

We now expand on how the right-hand side simplifies to 1 in the commuting case. First, the $c-1$ commutator-dependent terms vanish. For example, consider the commutator 
$\left[e^{- \beta_{c-1}^\bs Q_{c-1}^\text{tot}},  \, 
e^{- \Delta \beta_c Q_c^\as} 
U
e^{\Delta \beta_c Q_c^\as}\right] $.
The first argument commutes with $e^{\pm \Delta \beta_c Q_c^\as} $ because the charges commute. Also, the first argument commutes with $U$ because of charge conservation. Hence the commutator vanishes. 

Second, the first term in Eq.~\eqref{eq_chrg_ft_help1} equals 1. The thermal state [Eq.~\eqref{eq_GGE}] expands as a product so the term simplifies as:
\begin{align}
    \Tr( U^\dagger e^{-\Delta \beta_1 Q_1^{\as} } \ldots e^{-\Delta \beta_c Q_c^{\as}}  U e^{\Delta \beta_c Q_c^{\as}} \ldots e^{\Delta \beta_1 Q_1^{\as}} \rho )
   % % %
   &= \Tr( U^\dagger \prod_\gamma e^{-\Delta \beta_\gamma Q_\gamma^{\as} } U \prod_\delta e^{\Delta \beta_\delta Q_\delta^{\as}} \frac{1}{Z} \prod_\alpha 
   e^{-\beta^{\tt A}_\alpha Q_\alpha^{\tt A} - \beta^{\tt B}_\alpha Q_\alpha^{\tt B} } ) \\
   & = \Tr( U^\dagger \prod_\gamma e^{-\Delta \beta_\gamma Q_\gamma^{\as} } U \frac{1}{Z} \prod_\alpha e^{-   
   \beta^{\tt B}_\alpha Q_\alpha^{\text{tot}} } )
   \\
   =1.
\end{align}
The third equality follows from charge conservation.

\subsection{Detailed charge conservation}
\label{app:detailed}

The KDQ obeys detailed charge conservation if $\tilde{p}_{\rm F}$ is nonzero only when evaluated on indices that satisfy
\begin{equation}
\lambda_{\alpha, i^\as_\alpha}+\lambda_{\alpha, i^\bs_\alpha}=\lambda_{\alpha, f_\alpha^\as}+\lambda_{\alpha, f_\alpha^\bs}, \quad \forall \alpha .
\end{equation}
If the dynamics conserve just one charge, the KDQ obeys detailed charge conservation, we show in App.~\ref{app:single_charge_detailed}. Appendix~\ref{app:mult_charge_detailed} generalizes the argument to multiple commuting charges.

\subsubsection{Detailed charge conservation in the presence of only one charge}
\label{app:single_charge_detailed}

Here, we prove that the KDQ obeys detailed charge conservation in the presence of only one charge. We omit the charge's index to simplify notation.
The total charge eigendecomposes as $Q^\as+Q^\bs= \sum_k \lambda_k \Pi_k$.
The eigenprojectors decompose as
$\Pi_k 
   = \sum_{i^\as, i^\bs \, : \, 
   \lambda_{i^\as}+\lambda_{i^\bs}=\lambda_k} 
   \Pi_{i^\as} \otimes \Pi _{i^\bs}$.
Since $U$ commutes with $Q$, $U$
has the same block-diagonal structure: 
$U=\sum_k B_k$, wherein $B_k$ has support only on the subspace projected onto by $\Pi_k$. 

The relevant KDQ is
$\tilde{p}_{\rm F}(i^\as, i^\bs ; f^\as ,f^\bs)
=\Tr( U^\dagger \Pi_f U \Pi_i \rho)$.
By the projectors' orthogonality, 
$\Pi_f U = \Pi_f B_{k}$, wherein $k$ satisfies $\lambda_k=\lambda_{f^\as} +\lambda_{f^\bs}$. By the same logic,
$B_{k} \Pi_i=0$, unless $\lambda_{k}=\lambda_{i^\as} + \lambda_{ i^\bs}$. Hence $\Pi_{f^\as f^\bs} U \Pi_i$ and so 
$\tilde{p}(i^\as, i^\bs ; f^\as ,f^\bs)$ is nonzero only if 
$\lambda_{f^\as} + \lambda_{ f^\bs}
=\lambda_{k}=\lambda_{i^\as} + \lambda_{ i^\bs}$, 
satisfying detailed charge conservation.

\subsubsection{Detailed charge conservation in the presence of commuting charges}
\label{app:mult_charge_detailed}

Suppose that the dynamics conserve multiple charges that commute with each other. We show that the KDQ satisfies detailed charge conservation. 
Recall the form of $\tilde{p}_{\rm F}$ in Eq.~\eqref{KDdef}.
Since the charges commute, every eigenprojector in $\tilde{p}_{\rm F}$ commutes with every other.
Thus, rearranging the projectors does not alter the quasiprobability. 
We bring the initial and final charge-$\alpha$ eigenprojectors
beside $U$:
\begin{align}
   \tilde{p}_{\rm F}(i^\as_1,i^\bs_1, \ldots, i^\as_c,i^\bs_c \; ; \; f^\as_c,f^\bs_c, \ldots, f^\as_1,f^\bs_1)
   &=  \Tr( U^\dagger 
   \left[ \Pi_{1, f_1 } \ldots 
          \Pi_{c, f_c } \right]   
   U 
   \left[ \Pi_{c, i_c } \ldots \Pi_{1, i_1 } \right] 
   \rho ) \\ 
   & = \Tr( U^\dagger \ldots \Pi_{\alpha, f_\alpha }   U \Pi_{\alpha, i_\alpha } \ldots \rho ).
\end{align}
By the reasoning for one charge,
unless $\lambda_{\alpha, i^\as_\alpha}+\lambda_{\alpha, i_\alpha^\bs}
=\lambda_{\alpha, f_\alpha^\as}+\lambda_{\alpha, f_\alpha^\bs}$,
the projected unitary
$ \Pi_{\alpha, f_\alpha }   U \Pi_{\alpha, i_\alpha } =0$, and hence $\tilde{p}_{\rm F} =0$. 
This conclusion governs an arbitrary $\alpha$, so $\tilde{p}_{\rm F}$ satisfies detailed charge conservation.

\subsubsection{Fluctuation theorem's connection to nonconserving trajectories}
\label{ftandnoncons}

We now show how the two terms in the $\schrg$ fluctuation theorem [Eq.~\eqref{noncommft_app}] are related to (non)conserving trajectories. 
According to App.~\ref{app:schrg_FT}, the first term on the fluctuation theorem's RHS is $\expval{e^{ -\kappa }}$, wherein $\kappa\coloneqq \sum_{\alpha=1}^c \Delta \beta_\alpha ( \lambda_{\alpha, f_\alpha^\as} - \lambda_{\alpha, i_\alpha^\as} )$ [Eq.~\eqref{chrgftfirsterm}]. Since the average decomposes as $\expval{\cdot}=\expval{\cdot}_\text{cons}+\expval{\cdot}_\text{noncons}$, the fluctuation theorem's first term can contain contributions from conserving and nonconserving trajectories.

To identify the second term's relation to nonconserving trajectories, we rewrite the fluctuation theorem:
\begin{align}
    \expval{e^{-\schrg}}&= \expval{e^{-\kappa}} + \expval{e^{-\schrg}-e^{-\kappa}}\\
    &= \expval{e^{-\kappa}} + \expval{e^{-\schrg}-e^{-\kappa}}_\text{cons}+\expval{e^{-\schrg}-e^{-\kappa}}_\text{noncons}\\
    &= \expval{e^{-\kappa}} + \expval{e^{-\schrg}-e^{-\kappa}}_\text{noncons} \, .
\end{align}
The third line follows because $\schrg=\kappa$ on conserving trajectories, by the $\schrg$ definition~\eqref{schrg} and the definition~\eqref{dcc} of conserving trajectories.
Therefore, the fluctuation theorem's second term equals the average, over nonconserving trajectories, of $e^{-\schrg}-e^{-\kappa}$.

\section{Surprisal stochastic entropy production}
\label{surpapp}

This appendix supports claims made about $\ssurp$ in Sec.~\ref{sec_surp_SEP}. First, we complete the motivation for $\ssurp$'s definition (App.~\ref{app_surp_motivate}).
We show that the charge and surprisal formulae equal each other in the commuting case (App.~\ref{app:surp_chrg_agree}); 
calculate $\langle \ssurp \rangle$, proving Eq.~\eqref{avgs} (App.~\ref{app:surp_avg}); and prove the fluctuation theorem~\eqref{surpft} for $\ssurp$ (App.~\ref{surpftapp}).

\subsection{Motivation for the suprisal SEP formula}
\label{app_surp_motivate}
Section~\ref{sec_surp_SEP} partially motivated the $\ssurp$ definition~\eqref{ssurp}. Information theory suggested that the SEP depend on surprisals. The surprisals are of particular probabilities. Why those probabilities? This appendix motivates the choice.

We build on the first paragraph in Sec.~\ref{sec_traj_def}. For simplicity, we suppose that only energy is ever measured. A classical thermodynamic story motivated the conventional SEP formula,
\begin{align}
   \label{eq_straj_app}
   \log \left( 
   p_{\rm F} ( E^{\as}_i, E^{\bs}_i \, ; \, E^{\as}_f, E^{\bs}_f ) /
   p_{\rm R}( E^{\as}_f, E^{\bs}_f \, ; \, 
E^{\as}_i, E^{\bs}_i ) 
   \right) .
\end{align}
The $p_{\rm F}$ denotes the probability of observing the forward trajectory---the probability of observing $E^{\as}_i$ and $E^{\bs}_i$, times the probability of observing $E^{\as}_f$ and $E^{\bs}_f$, conditioned on the initial observations and on the forward dynamics:
\begin{align}
   & p_{\rm F} ( E^{\as}_i, E^{\bs}_i \, ; \, E^{\as}_f, E^{\bs}_f )
   = p ( E^{\as}_i, E^{\bs}_i ) \times
   p ( E^{\as}_f, E^{\bs}_f \, | \,
   E^{\as}_i, E^{\bs}_i, \text{forward dynamics} ) .
\end{align}
The reverse probability decomposes analogously:
\begin{align}
   & p_{\rm R}( E^{\as}_f, E^{\bs}_f  \, ; \, 
   E^{\as}_i, E^{\bs}_i )
   = p( E^{\as}_f, E^{\bs}_f, ) \times
   p( E^{\as}_i, E^{\bs}_i, \, | \,
   E^{\as}_f, E^{\bs}_f, \text{reverse dynamics} ) .
\end{align}
The dynamics are reversible, so the conditional probabilities cancel in~\eqref{eq_straj_app}~\cite{Landi2021}. The trajectory SEP formula reduces to
    $\log \left( 
   p( E^{\as}_i, E^{\bs}_i ) /
   p( E^{\as}_f, E^{\bs}_f ) \right) .$
This SEP formula is the difference of two surprisals. Each is evaluated on the probability of, upon preparing each system and measuring its energy, observing some outcome. The analog, in the noncommuting case, is Eq.~\eqref{ssurp}.

\subsection{Equivalence of \texorpdfstring{$\ssurp$}{TEXT} and \texorpdfstring{$\schrg$}{TEXT} in the commuting case}
\label{app:surp_chrg_agree}

Throughout this appendix, we assume that the charges commute with each other: $[Q_\alpha, Q_{\alpha'}] = 0$ $\forall \alpha, \alpha'$.
We establish that $\ssurp$ and $\schrg$ agree: $\ssurp=\schrg$ \emph{when evaluated on any trajectory for which $\tilde{p}_{\rm F}\neq0$}. The restriction on trajectories is a technicality; on a trajectory that never occurs, the SEPs' values are irrelevant to any calculation.

For convenience, we assume that charges' eigenvalues are ordered as follows. $Q_1$'s eigenvalues are ordered arbitrarily. Each other $Q_\alpha$'s eigenvalues are ordered so that the $j^\th$ eigenvalue, $\lambda_{\alpha, j}$, corresponds to the same eigenvalue as $\lambda_{1, j}$. This ordering is possible due to the charges' nondegeneracy: They all have the same number of eigenspaces.

$\tilde{p}_{\rm F}\neq 0$ only on trajectories in which all initial indices equal each other, as do all final indices: $i_1 = i_2 = \ldots = i_c $, and $f_1 = f_2 = \ldots = f_c$. This claim follows from the ordering stipulated in the last paragraph:
Up to a relabeling of eigenspaces, $\Pi_{\alpha, i_\alpha} \Pi_{\zeta, i_{\zeta}} = 0$, unless $i_\alpha=i_\zeta$. 
The claim now follows immediately
from the KDQ's definition [Eq.~\eqref{KDdef}],
    \begin{align}
    \label{KDdef_app} &
    \tilde{p}_{\rm F} (i_1, i_2, \ldots, i_c \, ; \,
    f_c, f_{c-1}, \ldots, f_1)
    \coloneqq
    \Tr \left( U^\dag
    \left[ \Pi_{1, f_1} \Pi_{2, f_2} \ldots \Pi_{c, f_c} \right]
    U 
    \left[ \Pi_{c, i_c} \ldots \Pi_{2,i_2} \Pi_{1,i_1}  \right]
    \rho \right) .
\end{align}
On the trajectories on which $\tilde{p}_{\rm F}\neq0$, we can simplify the $\schrg$ formula [Eq.~\eqref{schrg}] by equating 
all the charges' $i^\as$ indices, equating the charges' $i^\bs$ indices, equating the charges' $f^\as$ indices, and equating the charges' $f^\bs$ indices:
\begin{equation}
   \schrg 
   = \sum_{\zeta=1}^c \left[ 
   \beta^\as_{\zeta} \left( \lambda_{\zeta, f^\as_\alpha} - \lambda_{\zeta, i^\as_\alpha} \right) 
   + \beta^\bs_{\zeta} \left( \lambda_{\zeta, f^\bs_\alpha} 
   - \lambda_{\zeta, i^\bs_\alpha} \right)
   \right],   
   \; \; \text{if} \; \; \tilde{p}_{\rm F} \neq 0. \label{kronschrg}
\end{equation}

Consider now the surprisal SEP, which is defined in terms of measurement probabilities [Eq.~\eqref{ssurp}]. These probabilities simplify in the commuting case. Consider strongly measuring the shared eigenbasis of of system $\xs$'s charges. Outcome $i_\alpha^\xs$ obtains with a probability 
  \begin{align}
  \Tr ( \Pi_{\alpha,i_\alpha^\xs}^\xs \rho^\xs )
  &=  \frac{1}{Z^\xs} \Tr ( \Pi_{\alpha,i_\alpha^\xs}^\xs  
  e^{-\sum_{\zeta=1}^c \beta_\zeta^\xs Q^\xs_\zeta} )
  =  \frac{1}{Z^\xs} e^{- \sum_\zeta  \beta_\zeta^\xs \lambda_{\zeta,i^\xs_\alpha} } \Tr ( \Pi_{\alpha,i_\alpha^\as}^\as ) 
  =  \frac{1}{Z^\xs} e^{- \sum_\zeta  \beta_\zeta^\xs \lambda_{\zeta,i^\xs_\alpha} } \, .
\end{align}
The last equality follows from the charges' nondegeneracy: The eigenprojectors are rank-one. The choice of $\alpha$ is arbitrary, because all the charges have the same product bases. 
The surprisal SEP is therefore
\begin{align}
    \ssurp & = \log( 
    \frac{e^{ -\sum_{\zeta } \beta_\zeta^\as \lambda_{\zeta, i^\as_\alpha}} \
    e^{-\sum_{\zeta } \beta_{\zeta }^\as \lambda_{\zeta, i^\as_\alpha }}}
    { e^{ -\sum_{\zeta } \beta_\zeta^\as \lambda_{\zeta, f^\as_\alpha}} \
    e^ { -\sum_{\zeta} \beta_{\zeta }^\as \lambda_{\zeta, f^\as_\alpha }}} )
   \\
    &=\sum_{\zeta} \left[ 
   \beta^\as_{\zeta} \left( \lambda_{\zeta, f^\as_\alpha} - \lambda_{\zeta, i^\as_\alpha} \right) 
   + \beta^\bs_{\zeta} \left( \lambda_{\zeta, f^\bs_\alpha} 
   - \lambda_{\zeta, i^\bs_\alpha} \right)
   \right] 
    \, ,
\end{align}
in agreement with Eq.~\eqref{kronschrg}.

\subsection{Calculation of  \texorpdfstring{$\langle \ssurp \rangle$}{TEXT}}
\label{app:surp_avg}
Here, we prove Eq.~\eqref{avgs},
\begin{equation}
   \label{avgs_app}
   \hspace{-0.6cm} \expval{\ssurp} =   
   D \LParen \rho_{\rm f} || \Phi_\alpha \left( \rho \right) \RParen 
   - D \LParen \rho || \Phi_\alpha(\rho) \RParen \, . 
\end{equation}
$\ssurp$ [Eq.~\eqref{ssurp}] depends implicitly on fewer indices than $\tilde{p}_{\rm F}$ depends on [Eq.~\eqref{KDdef}].
Therefore, some of the indices disappear (yield factors of unity) when summed over in $\expval{\ssurp}$.
Let us substitute $\ssurp$'s definition into the LHS of~\eqref{avgs_app}, then substitute in each probability's definition. Finally, we rearrange terms:
\begin{align}
\nonumber
\langle \ssurp \rangle
    & = \left\langle 
    \log( 
   \frac{p_\alpha(i^\as_\alpha, i^\bs_\alpha) }{p_\alpha(f^\as_\alpha,f^\bs_\alpha) } )
   \right \rangle
  = 
  \sum_{i_\alpha,  f_\alpha}
  \Tr( U^\dagger \Pi_{\alpha, f_\alpha }  U \Pi_{\alpha, i_\alpha } \rho )  
  \log( \frac{\Tr(\Pi_{\alpha, i_\alpha} \rho) 
  }{\Tr(\Pi_{\alpha, f_\alpha} \rho ) } )
  \\
  & = \sum_{i_\alpha}
  \Tr(  \Pi_{\alpha, i_\alpha } \rho )  
  \log \LParen 
  \Tr(\Pi_{\alpha, i_\alpha} \rho) \RParen
  - \sum_{f_\alpha}
  \Tr(  \Pi_{\alpha, f_\alpha }  \rho_{\rm f})  
  \log \LParen
  \Tr(\Pi_{\alpha, f_\alpha} \rho ) \RParen 
  \\
  & = \Tr( \left[ \sum_{i_\alpha} \log \LParen 
  \Tr(\Pi_{\alpha, i_\alpha} \rho) \RParen
    \Pi_{\alpha, i_\alpha }\right] \rho )  
  - \Tr( \left[\sum_{f_\alpha}
    \Pi_{\alpha, f_\alpha } \log \LParen
  \Tr(\Pi_{\alpha, f_\alpha} \rho ) \RParen\right] \rho_{\rm f})     .\label{eq:avg_surp_bipartite}
\end{align}
To simplify the traces, we recall the channel that dephases $\rho$ with respect to $Q_\alpha$'s product basis:
$\Phi_\alpha(\rho) \coloneqq \sum_{i_\alpha} \Pi_{\alpha, i_\alpha}\,  \rho \,  \Pi_{\alpha, i_\alpha} = \sum_{i_\alpha} 
\Tr(\Pi_{\alpha, i_\alpha} \, \rho) \, 
\Pi_{\alpha, i_\alpha} $.
In terms of this channel,
\begin{equation}\label{eq:log_dephase}
    \log\LParen \Phi_\alpha(\rho) \RParen = 
\sum_{i_\alpha} 
\log\LParen \Tr(\Pi_{\alpha, i_\alpha} \, \rho) 
\RParen \,
\Pi_{\alpha, i_\alpha}.
\end{equation} 
Substituting into ~\eqref{eq:avg_surp_bipartite} yields
$\langle \ssurp \rangle 
= \Tr(\rho \log \LParen \Phi_\alpha(\rho) \RParen )
- \Tr(\rho_{\rm f} \log \LParen \Phi_\alpha(\rho) \RParen )$.
This expression can be rewritten in terms of relative entropies. 
By the von Neumann entropy's unitary invariance,
$0 = S_{\rm vN}(\rho) - S_{\rm vN}(\rho_{\rm f}) = \Tr\LParen \rho_{\rm f} \log(\rho_{\rm f})\RParen - \Tr\LParen\rho \log(\rho)\RParen$.
Adding this $0$ onto our $\langle \ssurp \rangle$ expression yields Eq.~\eqref{avgs_app}.

\subsection{Proof of fluctuation theorem for \texorpdfstring{$ \ssurp$}{TEXT}}
\label{surpftapp}

Here, we prove the fluctuation theorem~\eqref{surpft},
\begin{align}
   \label{surpft_app}
   \expval{e^{-\ssurp}} 
   =  1+\Tr( U^\dagger   \Phi_\alpha(\rho)   U \Delta \rho_\alpha \rho ) .
\end{align}
Also, we present the correction's form.

Recall the definition of the coherent difference $\Delta \rho_\alpha =  \Phi_\alpha(\rho)^{-1} -\rho^{-1} $. 
We substitute into $\expval{e^{-\ssurp}}$ the definitions of $\ssurp$ [Eq.~\eqref{ssurp}] and $\tilde{p}_{\rm F}$ [Eq.~\eqref{KDdef}]. As with the derivation of $\expval{\ssurp}$, some of the indices are marginalized over.
Rearranging factors:
\begin{align}
 \expval{e^{-\ssurp}} 
    & =  
    \sum_{ i_1^\as , i_2^\as ,\ldots, i_c^\as,\;  f_1^\as , f_2^\as ,\ldots, f_c^\as }
    \Tr( U^\dagger  \Pi_{1, f_1}  \ldots 
    \Pi_{c, f_c}  U  \Pi_{c, i_c}  \ldots 
    \Pi_{1, i_1}  \rho ) 
    \left( 
    \frac{ p_\alpha (f^\as_\alpha,f^\bs_\alpha)} {
    p_\alpha(i^\as_\alpha, i^\bs_\alpha)} \right) \\
   & = \sum_{i_\alpha^\as, i_\alpha^\bs, \; \; f_\alpha^\as, f_\alpha^\bs} 
   \Tr \left( U^\dagger  
   \left[ \Pi_{\alpha, f^\as_\alpha}^\as \otimes
   \Pi_{\alpha, f^\bs_\alpha }^\bs \right] U 
   \left[ \Pi_{\alpha, i^\as_\alpha}^\as \otimes 
          \Pi_{\alpha, i^\bs_\alpha }^\bs \right] 
   \rho \right) 
   \left( \frac{p_\alpha(f^\as_\alpha,f^\bs_\alpha)} {p_\alpha(i^\as_\alpha, i^\bs_\alpha)} \right) \\
   \label{eq_surp_FT_help1}
   & =  \Tr \left( U^\dagger
   \left\{
   \sum_{ f_\alpha^\as, f_\alpha^\bs}
   \left[ \Pi_{\alpha, f^\as_\alpha}^\as \otimes \Pi_{\alpha, f^\bs_\alpha }^\bs \right] \
   p_\alpha(f^\as_\alpha,f^\bs_\alpha)
   \right\}
   U 
   \left\{ \sum_{i_\alpha^\as, i_\alpha^\bs}
   \left[ \Pi_{\alpha, i^\as_\alpha}^\as \otimes \Pi_{\alpha, i^\bs_\alpha }^\bs \right] \
   p_\alpha(i^\as_\alpha, i^\bs_\alpha)^{-1}
   \right\}  \rho \right) \, . 
\end{align}
The summands simplify to dephased states and inverses thereof: 
$\sum_{ f_\alpha^\as, f_\alpha^\bs}
    (\Pi_{\alpha, f^\as_\alpha}^\as \otimes \Pi_{\alpha, f^\bs_\alpha }^\bs) \
    p_\alpha(f^\as_\alpha,f^\bs_\alpha)=\Phi_\alpha(\rho)$,
and 
$\sum_{i_\alpha^\as, i_\alpha^\bs}
    (\Pi_{\alpha, i^\as_\alpha}^\as \otimes \Pi_{\alpha, i^\bs_\alpha }^\bs) \
     p_\alpha(i^\as_\alpha, i^\bs_\alpha)^{-1}= \Phi_\alpha(\rho)^{-1}=\rho^{-1} + \Delta \rho_\alpha$ \,.
We substitute into Eq.~\eqref{eq_surp_FT_help1}, multiply out, and simplify:
\begin{align}
   \expval{e^{-\ssurp}} 
   &= \Tr \LParen U^\dagger  
   \Phi_\alpha(\rho) 
   U \left[ \rho^{-1} + \Delta \rho_\alpha \right] \rho \RParen 
   = \Tr \LParen U^\dagger  \Phi_\alpha(\rho)   U \rho^{-1} \rho \RParen  
   + \Tr \LParen U^\dagger   \Phi_\alpha(\rho)   U \Delta \rho_\alpha \rho \RParen \\
   &= 1 
   + \Tr( U^\dagger   \Phi_\alpha(\rho)   U \Delta \rho_\alpha \rho ) \, . 
\end{align}
The second term is the correction sourced by coherences. It vanishes if $\Delta \rho_\alpha=0$.

\section{Trajectory stochastic entropy production}
\label{trajapp}

This appendix concerns the trajectory SEP formula (Sec.~\ref{sec_traj_def}). 
We further motivate our choice of reverse quasiprobability, $\tilde{p}_{\rm R}$, in App.~\ref{app_traj_rev_def}.
We prove that $\straj = \schrg$ in the commuting case (Sec.~\ref{sec_traj_def}) in App.~\ref{app:detailed}.
In App.~\ref{app:traj_avg_pure_states}, we calculate $\langle \straj\rangle$ [prove Eq.~\eqref{avgt}].

\subsection{Choice of reverse quasiprobability}
\label{app_traj_rev_def}

Section~\ref{KD} introduced the forward quasiprobability [Eq.~\ref{KDdef}],
\begin{align}
   \label{eq_traj_rev_def_help1} &
   \tilde{p}_{\rm F} (i_1, i_2, \ldots, i_c \, ; \,
   f_c, f_{c-1}, \ldots, f_1)
   \coloneqq \Tr \left( U^\dag
   \left[ \Pi_{1, f_1} \Pi_{2, f_2} \ldots \Pi_{c, f_c} \right]
   U 
   \left[ \Pi_{c, i_c} \ldots \Pi_{2,i_2} \Pi_{1,i_1}  \right]
   \rho \right),
\end{align}
and Sec.~\ref{sec_traj_def} introduced the reverse quasiprobability, 
\begin{align}
\label{eq_traj_rev_def_help2}
    \tilde{p}_{\rm R}
    (f_1, f_2, \ldots, f_c \, ; \, 
    i_c, i_{c-1}, \ldots, i_1)
\coloneqq \Tr \left( 
   \left[ \Pi_{1, f_1} \Pi_{2, f_2} \ldots \Pi_{c, f_c} \right]
   U 
   \left[ \Pi_{c, i_c} \ldots \Pi_{2,i_2} \Pi_{1,i_1}  \right]
   U^\dag
   \rho \right)
    \, .
\end{align}
We motivated $\tilde{p}_{\rm R}$'s definition operationally in Sec.~\ref{sec_traj_def}. However, other possible definitions could satisfy the constraint that, in the commuting case, $\straj$ must reduce to $\schrg$ and $\ssurp$. 
Prior literature features nonunique reverse distributions (of probabilities, rather than quasiprobabilities)~\cite{Manzano2018, Landi2021}.
The authors there distinguished one reverse distribution by specifying one of multiple possible reverse protocols.
Similarly, we specify a reverse protocol from which $\tilde{p}_{\rm R}$ can be inferred experimentally.
We therefore regard $\tilde{p}_{\rm R}$ as capturing the notion of time reversal.

We can associate each KDQ with a protocol---the key stage of an experiment used to infer the quasiprobability (see App.~\ref{app_Motivate_KDQ} and~\cite{NYH_18_Quasiprobability}).
One begins at one end of the trace's argument and proceeds toward the opposite end. If a projector appears on just one side of $\rho$, that projector corresponds to a weak measurement. The $\tilde{p}_\text{R}$ protocol is a time reverse of the $\tilde{p}_\text{F}$ protocol, we show.

We associate $\tilde{p}_\text{F}$ with a protocol by reading the indices in~\eqref{eq_traj_rev_def_help1} from right to left:
One measures charges $1$ through $c$ weakly, evolves the state forward in time, and then measures charges $c$ through $1$ weakly.
Before interpreting $\tilde{p}_\text{R}$, 
we cycle $\rho$ to the trace's LHS: 
\begin{align}
   \tilde{p}_{\rm R} (f_1, f_2, \ldots, f_c \, ; \, 
    i_c, i_{c-1}, \ldots, i_1)
   \coloneqq \Tr \left(   \rho 
   \left[ \Pi_{1,f_1} \ldots \Pi_{c, f_c} \right] 
   U \left[ \Pi_{c, i_c} \ldots \Pi_{1, i_1} \right] U^\dagger \right) .
\end{align}
Then, we read from left to right. In the associated protocol, one measures charges $1$ through $c$ weakly, evolves the state backward in time (under $U^\dag$), and then measures charges $c$ through $1$ weakly. 
Relative to the forward protocol, the time evolution is reversed, as is the list of measurement outcomes.

\subsection{Equivalence of \texorpdfstring{$\straj$}{TEXT}  and   \texorpdfstring{$\ssurp$}{TEXT} in the commuting case}
\label{app_traj_reduce}
In this appendix, we show  that $\straj = \ssurp$ in the commuting case.
Equation~\eqref{eq:straj_single_charge} defines the trajectory SEP:
\begin{align}
   \straj &=\log \left(
   \bra{i_1} \rho U^\dagger \ket{f_1} /
   \bra{i_1} U^\dagger \rho \ket{f_1} \right) .
   \label{eq:straj_single_charge_app}
\end{align}
Since the initial state is diagonal with respect to the $1^{\rm{st}}$ product basis,
$\bra{i_1}\rho 
   =  \Tr(\Pi_{i_1} \rho) \bra{i_1}$,
and
   $\rho \ket{f_1}
   = \Tr(\Pi_{f_1} \rho) \ket{f_1} .$
Therefore,
\begin{align}
    \straj &= \log \left( \frac{ \Tr(\Pi_{i_1} \rho)
   \bra{i_1} U^\dagger \ket{f_1}}{
    \Tr(\Pi_{f_1} \rho) \bra{i_1} U^\dagger \ket{f_1} } \right)
    = \log \left( \frac{ \Tr(\Pi_{i_1} \rho)}{\Tr(\Pi_{f_1} \rho) } \right).
\end{align}
The final expression is $\ssurp$ [Eq.~\eqref{ssurp}] with $\alpha=1$. (The choice of $\alpha$ is irrelevant in the commuting case, since all the product bases coincide.)
We have already established the equivalence of $\ssurp$ and $\schrg$ in the commuting case (App.~\ref{app:surp_chrg_agree}). Therefore, we have shown that all three SEP formulae agree when charges commute.

\subsection{Calculation of \texorpdfstring{$\langle \straj \rangle$}{TEXT} when the initial state \texorpdfstring{$\rho$}{TEXT} is pure}
\label{app:traj_avg_pure_states}

Let the initial state in Sec.~\ref{bgsetup} be pure:
$\rho = |\psi\rangle \langle \psi |$.
Under this assumption, we derive Eq.~\eqref{avgt}:
\begin{align}
   \label{avgt_app}
   \expval{\straj} &= \tfrac{1}{2} \big[ D\boldsymbol{(}\rho || \Phi^1(U^\dagger \rho U)\boldsymbol{)} + D\boldsymbol{(}\rho||  U^\dagger \Phi^1(\rho) U\boldsymbol{)} 
   - D\boldsymbol{(}\rho || \Phi^1(\rho)\boldsymbol{)}-D\boldsymbol{(}\rho_\text{f} || \Phi^1(\rho_\text{f})\boldsymbol{)}   \big]  
   + i\langle \phi_\text{F} - \phi_\text{R}\rangle \, .
\end{align}
Our proof consists of three steps. First, we express $\straj$ in terms of weak values,
slightly rewriting Eq.~\eqref{eq:straj_cartesian}:
\begin{align}
   \label{eq:straj_cartesian_app}
   \straj 
      & = \log \left( \frac{ 
   \big\lvert 
   {}_{f_1}\langle \Pi_{1,i_1} \rangle_{\rho} \big\rvert }{
  \big\lvert 
  {}_{i_1}\langle \Pi_{1, f_1} 
   \rangle_{\rho}  \big\rvert }  \,
   \frac{ \Tr \big( \Pi''_{1,f_1} \rho \big) }{
          \Tr \big( \Pi_{1, i_1} \rho \big) }
   \right)
   + i(\phi_\text{F} - \phi_\text{R}) .
\end{align}
Second, we prove that $i\langle \phi_{\rm F} - \phi_{\rm R}\rangle$ is imaginary. 
Finally, we prove that the logarithm's average is real and equals the relative-entropy sum in Eq.~\eqref{avgt_app}. By implication, 
$i\langle \phi_{\rm F} - \phi_{\rm R}\rangle
= \Im( \langle\straj\rangle )$.
This result supports the main text's claim that 
$\Im( \langle\straj\rangle ) \neq 0$ signals contextuality.

Let us show that 
$i\langle \phi_\text{F} - \phi_\text{R}\rangle$
is imaginary.
Let $z$ denote any complex number. We denote its argument's (its complex phase's) principal value by ${\rm Arg}(z)$ and the multivalued argument function by ${\rm arg}(z)$.
$\phi_{\rm F} - \phi_{\rm R}$ equals an argument by Eq.~\eqref{eq:straj_single_charge} and the polar forms in Sec.~\ref{sec_traj_def}:
For each index pair $(i_1, f_1)$, there is a branch of
$\rm{arg}$ such that
\begin{align}
    \phi_{\rm F} - \phi_{\rm R} 
   & = {\rm arg}\left(
   \frac{\langle i_1|\psi \rangle \langle \psi | U^\dagger | f_1\rangle}{\langle i_1 | U^\dagger | \psi \rangle \langle \psi | f_1\rangle} 
   \right)
    = 
   {\rm Arg}\left( \langle i_1|\psi \rangle \right)
   + {\rm Arg}\left( \langle \psi | U^\dagger | f_1\rangle \right)
   - {\rm Arg}\left( \langle i_1 | U^\dagger | \psi \rangle \right)
   - {\rm Arg}\left( \langle \psi | f_1\rangle \right) \, .\label{eq:multiple_Args_app}
\end{align}
Each summand in Eq.~\eqref{eq:multiple_Args_app}'s RHS depends on one index, $i_1$ or $f_1$. (If $\rho$ is mixed, $\straj$ does not decompose in this manner. Hence new techniques are required to compute the average.)
Therefore, averaging $\phi_{\rm F} - \phi_{\rm R}$ with respect to $\tilde{p}_{\rm F}$ yields separate averages with respect to probability distributions. Hence $\langle \phi_{\rm F} - \phi_{\rm R}\rangle$ is real, and $i\langle \phi_{\rm F} - \phi_{\rm R}\rangle$ is imaginary.

Our task reduces to showing that 
\begin{align}\label{eq:real_avgt_app}
    \left\langle
    \log \Bigg( \frac{ 
   \big\lvert 
   {}_{f_1}\langle \Pi_{1,i_1} \rangle_{\rho} \big\rvert }{
  \big\lvert 
  {}_{i_1}\langle \Pi_{1, f_1} 
   \rangle_{\rho}  \big\rvert }  
   \frac{ \Tr \big(\Pi_{1, f_1}'' \rho \big)}{
          \Tr \big(\Pi_{1, i_1}' \rho \big) } \Bigg)
   \right\rangle
   = \tfrac{1}{2} \big[ 
   D\boldsymbol{(}\rho || \Phi^1(U^\dagger \rho U)\boldsymbol{)} 
   + D\boldsymbol{(}\rho||  U^\dagger \Phi^1(\rho) U\boldsymbol{)}
   - D\boldsymbol{(}\rho || \Phi^1(\rho)\boldsymbol{)}
   -D\boldsymbol{(}\rho_{\rm f} || \Phi^1(\rho_{\rm f})\boldsymbol{)}   \big].
\end{align}
The weak values have magnitudes of
\begin{equation}\label{eq:WV_mags_app}
    \big\lvert 
   {}_{f_1}\langle \Pi_{1,i_1} \rangle_{\rho} \big\rvert = \frac{\left[\Tr(\Pi_{1, f_1}'' \Pi_{1, i_1}\rho) \Tr(\rho \Pi_{1, i_1} \Pi_{1, f_1}'')\right]^{\frac{1}{2}}}{\Tr(\Pi_{1, f_1}'' \rho)}
    \quad {\rm and} \quad
    \big\lvert 
  {}_{i_1}\langle \Pi_{1, f_1} 
   \rangle_{\rho}  \big\rvert = \frac{\left[\Tr(\Pi_{1, i_1}' \Pi_{1, f_1}\rho)\Tr(\rho \Pi_{1, f_1} \Pi_{1, i_1}')\right]^{\frac{1}{2}}}{\Tr(\Pi_{1, i_1}'\rho)}\, .
\end{equation}
When we insert these expressions into Eq.~\eqref{eq:real_avgt_app}, the denominators in~\eqref{eq:WV_mags_app} cancel.
To simplify Eq.~\eqref{eq:real_avgt_app}'s LHS more,
we express $\rho$ and the rank-1 projectors as outer products: $\rho =|\psi\rangle \langle \psi|$, $\Pi_{1, i_1} = |i_1\rangle\langle i_1|$, etc.
Factors $\langle f_1| U | i_1\rangle$ and 
$\langle i_1 | U^\dagger | f_1 \rangle$ appear in the numerator and denominator. Canceling the factors, we compute Eq.~\eqref{eq:real_avgt_app}'s LHS: 
\begin{align}
    &\left\langle \log \left(
    \frac{\left[\Tr(\Pi_{1, f_1}'' \Pi_{1, i_1}\rho) \Tr(\rho \Pi_{1, i_1} \Pi_{1, f_1}'')\right]^{\frac{1}{2}}}
    {\left[ \Tr(\Pi_{1, i_1}' \Pi_{1, f_1}\rho)
    \Tr(\rho \Pi_{1, f_1} 
    \Pi_{1, i_1}')\right]^{\frac{1}{2}}}
    \right) \right\rangle
    = \left\langle\frac{1}{2}\log\left( \frac{\Tr( \Pi_{1, f_1} U\rho U^\dagger) \Tr( \Pi_{1, i_1} \rho)}{\Tr(\Pi_{1, f_1} \rho) \Tr(\Pi_{1, i_1} U^\dagger \rho U)}\right)\right\rangle\\
    &= \sum_{i_1, \ldots, i_c, \; f_1, \ldots, f_c} \frac{1}{2}\Tr(U^\dagger\Pi_{1, f_1} \ldots \Pi_{1, f_c}  U \Pi_{c, i_c} \ldots \Pi_{1, i_1} \rho)\log\left( \frac{\Tr( \Pi_{1, f_1} U\rho U^\dagger) \Tr( \Pi_{1, i_1} \rho)}{\Tr(\Pi_{1, f_1} \rho) \Tr(\Pi_{1, i_1} U^\dagger \rho U)}\right)\\
    &= \frac{1}{2} \Bigg[
    \sum_{f_1} \Tr( \Pi_{1, f_1} \rho_{\rm f}) 
    \log \LParen \Tr( \Pi_{1, f_1} \rho_{\rm f}) \RParen 
    + \sum_{i_1}\Tr( \Pi_{1, i_1} \rho) 
    \log \LParen \Tr( \Pi_{1, i_1} \rho) \RParen 
    \\ \nonumber &\quad \quad \quad 
    - \sum_{f_1} \Tr( \Pi_{1, f_1} \rho_{\rm f})
    \log \LParen \Tr( \Pi_{1, f_1} \rho) \RParen  
    -\sum_{i_1} \Tr(\Pi_{1, i_1}\rho) 
    \log \LParen \Tr(\Pi_{1, i_1} U^\dagger \rho U) \RParen
    \Bigg] \\
    &= \frac{1}{2}\left[
    \Tr \left( \rho_{\rm f} 
    \log \LParen \Phi_1(\rho_{\rm f}) \RParen \right)
    + \Tr\left( \rho  
    \log \LParen \Phi_1(\rho ) \RParen \right)
    -\Tr\left( \rho_{\rm f} 
    \log \LParen \Phi_1(\rho) \RParen \right)
    - \Tr\left( \rho  
    \log \LParen \Phi_1(U^\dagger \rho U ) \RParen \right)    \right].
    \label{eq:before_additional_dephase}
\end{align}
The final equality follows from Eq.~\eqref{eq:log_dephase}. 

By the von Neumann entropy's unitary invariance,
$\frac{1}{2}[S_{\rm vN}(\rho) - S_{\rm vN}(\rho) + S_{\rm vN}(\rho) - S_{\rm vN}(\rho_{\rm f})] = 0$. 
We add this zero to~\eqref{eq:before_additional_dephase}. 
By the trace's cyclicality,~\eqref{eq:before_additional_dephase}'s
$\Tr\big(\rho \log\LParen U^\dagger \Phi_1(\rho) U \RParen \big) = \Tr\big(U \rho U^\dagger \log\LParen \Phi_1(\rho) \RParen \big)$.
Rearranging terms yields the RHS of Eq.~\eqref{eq:real_avgt_app}. $\Re(\expval{\straj})$ is non-negative because it can be written as $\tfrac{1}{2} \big[ 
   D\boldsymbol{(} \Phi^1(\rho) || \Phi^1(U^\dagger \rho U)\boldsymbol{)} 
   + D\boldsymbol{(}  \Phi^1(\rho_{\rm f} ) ||\Phi^1(\rho)  \boldsymbol{)} \big]$.

When $\rho$ is pure, Eq.~\eqref{eq:multiple_Args_app} points, for each index pair $(i_1,f_1)$, to a choice of branch of the complex logarithm in the $\straj$ definition. This choice enables $\expval{\straj}$ to signal contextuality. If $\rho$ is mixed, in the absence of a more explicit $\expval{\straj}$ expression, it is not clear what choice enables $\expval{\straj}$ to signal contextuality.

\section{Symmetrized definitions}
\label{app_symmetrized}

In defining the KDQ and some SEPs, we choose projectors and the order in which to arrange them. 
Here, we discuss alternative KDQ and SEP definitions based on symmetrizing over these choices. We present the corresponding averages $\expval{\sigma}$. Via arguments similar to the ones for unsymmetrized $\sigma$'s, one can derive fluctuation theorems.

\subsection{Symmetrized Kirkwood-Dirac quasiprobability}
A primary motivation for using the KDQ is a desire to describe charges flowing in the absence of measurement disturbance.
Though based on weak measurements, the KDQ 
depends on the measurements' ordering, i.e., how the charges are labeled. 
But if the systems interact unitarily without being measured, 
relabeling the charges does not change any physical observables. This process's stochastic description, too, should respect this symmetry.
We take this for granted in the commuting case, in which
the measurements' ordering does not change the joint probability distribution. We can enforce this symmetry in the noncommuting case by averaging over all possible orderings of the initial projectors. (Once the initial projectors are ordered, the final projectors' ordering is fixed by the fourth sanity check in Sec.~\ref{sec_3formulae}.) We define the \emph{symmetrized KDQ} (SKDQ) as
\begin{equation}
\label{KDsymmdef}
	\tilde{p}_F^{\text{symm}} 
	\coloneqq \frac{1}{c!} \sum_{\tau\in S_c} 
	\Tr(U^\dagger \left[\Pi_{\tau(1), f_{\tau(1)}}\ldots \Pi_{\tau(c), f_{\tau(c)}} \right] U \left[\Pi_{\tau(c), i_{\tau(c)}} \ldots \Pi_{\tau(1), i_{\tau(1)}}\right] \rho).
\end{equation}
$S_c$ is the symmetric group of degree $c$, i.e., the set of all permutations of $\mathbbm{Z}_c$. 
$\tilde{p}_F^{\text{symm}}$ is invariant under every possible relabeling of the charges, respecting a fundamental symmetry of the undisturbed-charge-flow process. The symmetrized KDQ's single-index marginals are those of
$\tilde{p}_\text{F} $. In the commuting case, $\tilde{p}_F^{\text{symm}} = \tilde{p}_F$.

\subsection{Charge stochastic entropy production}

The charge SEP is already symmetric under every possible relabeling of the charges. Since the SKDQ has the same marginals as the KDQ, $\schrg$ has the same average:
\begin{equation}
    \expval{\schrg}_{{\text{symm}}}=\expval{\schrg}= \sum_\alpha \Delta \beta_\alpha \, \Delta \expval{Q_\alpha}.
\end{equation}

\subsection{Surprisal stochastic entropy production }
To obtain the symmetrized surprisal SEP, we average over the different bases: 
\begin{equation}
    \ssurp^{\text{symm}} \coloneqq \frac{1}{c} \sum_{\alpha=1}^c \log \left( 
   \frac{p_\alpha(i^\as_\alpha, i^\bs_\alpha) }{p_\alpha(f^\as_\alpha,f^\bs_\alpha) } \right) .
\end{equation}
The average now involves dephasing in all bases:
 \begin{equation}
     \expval{\ssurp^{\text{symm}}}_{{\text{symm}}} = \frac{1}{c} \sum_{\alpha=1}^c \left[ D \LParen \rho_{\rm f} || \Phi_\alpha \left( \rho \right) \RParen 
   - D \LParen \rho || \Phi_\alpha(\rho) \RParen \right].
 \end{equation}

\subsection{Trajectory stochastic entropy production }
A natural definition for the symmetrized trajectory SEP is
\begin{equation}\label{straj}
   \straj^{\text{symm}}=  \frac{1}{c} \left( \sum_{\alpha=1}^c \log \frac{
    \bra{i_\alpha} \rho U^\dagger \ket{f_\alpha}}{
    \bra{i_\alpha}U^\dagger \rho \ket{f_\alpha}} \right).
\end{equation}
This symmetrized SEP is closely related to the symmetrized KDQ. As in the unsymmetrized case, define the trajectory SEP as the ratio of forward and reverse quasiprobabilities. However, perform the averaging after taking the ratio:
\begin{align}
   \straj^{\text{symm}}
    &\coloneqq \frac{1}{c!} \sum_{\tau\in S_c} \log \left( \frac{\tilde{p}_{\rm F}(\tau)}{\tilde{p}_{\rm R}(\tau)} \right)
    = \frac{1}{c!} \sum_{\tau\in S_c}\log \left( \frac{
    \Tr(U^\dagger \left[\Pi_{\tau(1), f_{\tau(1)}}\ldots \Pi_{\tau(c), f_{\tau(c)}} \right] U \left[\Pi_{\tau(c), i_{\tau(c)}} \ldots \Pi_{\tau(1), i_{\tau(1)}}\right] \rho)}{ 
    \Tr(\left[\Pi_{\tau(1), f_{\tau(1)}}\ldots \Pi_{\tau(c), f_{\tau(c)}} \right] U \left[\Pi_{\tau(c), i_{\tau(c)}} \ldots \Pi_{\tau(1), i_{\tau(1)}}\right] U^\dagger \rho)} \right)\\ 
    &= \frac{1}{c} \sum_{\alpha=1}^c \log \left( \frac{
    \bra{i_\alpha} \rho U^\dagger \ket{f_\alpha}}{
    \bra{i_\alpha}U^\dagger \rho \ket{f_\alpha}} \right) \, .
\end{align}
The factors' cancellation simplifies the sum to be only over $c$ charges, rather than $c!$ permutations.
If $\rho$ is pure, the average is 
\begin{align}
   \label{avgt_app2}
   \expval{\straj} 
   & = \frac{1}{c} \sum_{\alpha=1}^c \bigg\{ \tfrac{1}{2} \big[ 
   D\boldsymbol{(}\rho || \Phi_\alpha( U^\dagger \rho U)\boldsymbol{)} 
   + D\boldsymbol{(}\rho||  U^\dagger \Phi_1(\rho) U\boldsymbol{)}  
   - D\boldsymbol{(}\rho || \Phi_1(\rho)\boldsymbol{)}
   - D\boldsymbol{(}\rho_{\rm f} || \Phi_1(\rho_{\rm f})\boldsymbol{)} \big]  
   + i \langle \phi_\text{F}^\alpha - \phi_\text{R}^\alpha\rangle_{\rm{symm}} \bigg\} .
\end{align}
The weak-value phases for general $\alpha$ are defined analogously to the weak-value phases in Sec.~\ref{sec_nonclass}.

\end{appendices}

\bibliographystyle{mod_revtex_titles}
\bibliography{mainbib}

\end{document}